\documentclass[useAMS,usenatbib,twocolumn]{mn2e}
\usepackage[dvips]{graphicx}
\usepackage{amsmath}
\usepackage{amsfonts}
\usepackage{amssymb}

\newcommand{\luna}{{\tt LUNA}}

\title[\luna: A transit algorithm for exomoons]{\luna: An algorithm for 
generating dynamic planet-moon transits}
\author[David M. Kipping]{David M. Kipping$^{1,2}$\thanks{E-mail:
dkipping@cfa.harvard.edu}\footnotemark[1]\\
$^{1}$Harvard-Smithsonian Center for Astrophysics, Garden St., Cambridge, 
MA 02138, USA \\
$^{2}$University College London, Dept. of Physics \& Astronomy, Gower St., 
London, WC1E 6BT, UK}
\begin{document}

\date{Accepted 2011 May 16. Received 2011 May 11; in original form 2011 February 
26}

\pagerange{\pageref{firstpage}--\pageref{lastpage}} \pubyear{2011}

\maketitle

\label{firstpage}


\begin{abstract}

It has been previously shown that moons of extrasolar planets may be detectable 
with the \emph{Kepler Mission}, for moon masses above $\sim0.2$\,$M_{\oplus}$ 
\citep{kipping:2009c}. Transit timing effects have been formerly identified as a 
potent tool to this end, exploiting the dynamics of the system. In this work, we 
explore the simulation of transit light curves of a planet plus a single moon 
including not only the transit timing effects but also the light curve signal 
of the moon itself. We introduce our new algorithm, \luna, which produces 
transit light curves for both bodies, analytically accounting for shadow 
overlaps, stellar limb darkening and planet-moon dynamical motion. By building 
the dynamics into the core of \luna, the routine automatically accounts for 
transit timing/duration variations and ingress/egress asymmetries for not 
only the planet, but also the moon.\\
We then generate some artificial data for two feasibly detectable hypothetical 
systems of interest: a i) prograde and ii) retrograde Earth-like moon around a 
habitable-zone Neptune for a M-dwarf system. 
We fit the hypothetical systems using \luna\ and demonstrate the feasibility of 
detecting these cases with \emph{Kepler} photometry.

\end{abstract}




\section{Introduction}
\label{sec:intro}

In recent years, the possibility of detecting the moons of extrasolar planets,
so-called ``exomoons'', has received increased attention \citep{sartoretti:1999,
han:2002,szabo:2006,simon:2009,lewis:2008,kipping:2009a,kipping:2009b,sato:2009,
weighing:2010}. Extrasolar moons may be frequent, temperate abodes for life and 
a determination of their prevalence would mould our understanding of the
abundance of life in the Universe. Although many techniques have been proposed, 
it is the transit method which seems to offer the greatest potential to detect 
habitable-zone moons in the near-future \citep{kipping:2009c}.

In general, there are two ways in which a moon can be identified using transits. 
The first of these is the moon's own transit light curve \citep{sartoretti:1999}
and the second is the family of techniques known as ``transit timing 
effects''. This includes both transit timing variations (TTV) 
\citep{sartoretti:1999} and transit duration variations (TDV). Further, TDV has 
two different components; one due to velocity variations (TDV-V) 
\citep{kipping:2009a} and one due to transit impact parameter variations 
(TDV-TIP) \citep{kipping:2009b}. TTV and TDV-TIP are both due to
the position of the planet varying in response to the moon's presence 
(positional wobble). In contrast, TDV-V is due to velocity variation of the host 
planet in response to the moon's presence (velocity wobble). Combining all of 
these effects allows for a unique solution (see \citeauthor{kipping:2009b} 
\citeyear{kipping:2009b} for details) for the moon-to-planet mass ratio and 
exomoon period (and therefore orbital semi-major axis through Kepler's Third Law).

To perform the observations, highly precise, continuous photometry is required
and the \emph{Kepler Mission} is thought to be the best instrument currently
available to conduct such a search \citep{kipping:2009c}.
So far, two searches for exomoons have been conducted using the \emph{Kepler} 
photometry, utilizing the transit timing techniques on Kepler-4b 
through 8b \citep{kippingbakos:2011a} and TrES-2b \citep{kippingbakos:2011b}. 
Despite null-results (which was expected as all of these planets are
hot-Jupiters), the studies indicate sensitivity into the sub-Earth mass regime, 
as predicted by \citet{kipping:2009c}. The transit timing equations are 
all built around the premise of a constant planet-star separation and constant 
planetary velocity during the timescale of the transit. Whilst this is 
generally a good approximation except for moons on very short periods, there is 
an obvious desideratum to make the expressions dynamic.

A possible problem with the current technique is that if the moon's light curve 
is detectable directly, then not only is this an extra piece of data we are 
ignoring but also it could fundamentally invalidate many of the transit timing 
methods. The ideal solution would be to simulate both the planet and moon 
transit light curves, including limb darkening which is critical for 
\emph{Kepler}. In addition, these light curves could be computed with the TTV 
and TDV effects inherently built into the model in a dynamic way. Not only 
would this improve upon the static approximations made in the TTV/TDV equations, 
but it would also bring in TTV/TDV of the moon itself.

Before constructing such an algorithm, one must keep the purpose of such a 
routine in mind. This algorithm will be used to search for exomoons which will 
be done by fitting light curves with the new algorithm. Given the large number 
of free parameters and inevitable intricate inter-parameter dependencies, a 
Monte Carlo based method seems required. Since such methods are inherently 
computationally expensive making often millions of calls to the simulation 
routine, then it is clear that any algorithm we design must be very fast to 
execute. This essentially excludes methods based upon pixelating the star or 
other numerical methods. The clear requirement then, is a completely analytic 
algorithm. The list of requirements are:

\begin{itemize}
\item[{\tiny$\blacksquare$}] Analytic (absolutely no numerical components)
\item[{\tiny$\blacksquare$}] Dynamic (inherently accounts for all timing 
effects)
\item[{\tiny$\blacksquare$}] Limb darkening incorporated (including non-linear 
laws)
\item[{\tiny$\blacksquare$}] All orbital elements accounted for (e.g. 
eccentricity, longitude of the ascending node, etc)
\end{itemize}

In summary, such an approach would not only offer many advantages over the 
previous timing methods, but would also be highly practical for conducting a
batch-style search for exomoons in archival data. In this paper, we introduce 
the fundamental equations needed to construct this algorithm, known as \luna, 
and apply it in some hypothetical examples.

In \S\ref{sec:lightcurves}, we describe the framework for
computing the transit light curves of a planet with a moon and outline the 
assumptions made. Accompanying details on the model used for the sky-projected 
motions of the planet and moon can be found in the Appendix 
(\S\ref{sec:motion}). In \S\ref{sec:areas}, we provide expressions for the 
eclipsing area of the moon in front of the star in all 27 possible principal 
configurations. In \S\ref{sec:examples}, example transits light curves are 
generated and re-fitted for habitable-zone exomoons detectable with 
\emph{Kepler}-class photometry. Finally, we discuss comparison to previous 
methods proposed in the literature and provide concluding remarks in 
\S\ref{sec:conclusions}.

\section{Light Curve Generation}
\label{sec:lightcurves}

\subsection{Converting Time to True Anomaly}
\label{sub:conversion}

To generate a light curve, the relative positions of the star, planet and moon
must be calculated at every time stamp in the photometric data set in question. 
In particular, one requires the sky-projected separation between the planet and 
star, $S_{P*}$, the moon and the star, $S_{S*}$ and the planet and the moon, 
$S_{PS}$. We direct the reader to the Appendix (\S\ref{sec:motion}) for details 
on the derivation of the expressions for these terms, where we employ an 
analytic approximation for the three-body problem which we dub as the nested 
two-body model \citep{weighing:2010}. The coordinate system employed throughout 
is also defined in the Appendix (\S\ref{sec:motion}). It can be shown that these 
three $S$ terms are a function of $f_{B*}$ and $f_{SB}$ which may be written, 
in general, as a function of time alone. Note that we define $f_{B*}$ as the 
true anomaly of the planet-moon barycentre around the star and $f_{SB}$ as the 
true anomaly of the satellite around the planet-moon barycentre. Therefore, a 
necessary prerequisite is to convert time stamps to true anomalies. Let us begin 
with considering the standard practice for a planet by itself.

\subsubsection{The Planet-Only Case}

The planet-only case is a very familiar one. However, we will cover the
conversion of time to true anomaly carefully. Writing out the steps explicity
will allow us to identify the necessary procedure for the more complicated
planet-moon case we will soon face.

Typically, we have a time series running from an initial time stamp of $t_1$ up 
to $t_N$ (where $N$ is the total number of measurements) with the time of 
transit minimum occurring at $\tau$. Our first task is to convert the time array 
into an $f$ array. To accomplish this we must define a reference point in
time at which the true anomaly is known. Whilst we are free to use any reference
point we so desire, a typical choice is the time of transit minimum, $\tau$. 
In the example above, this choice it will change our array to run from 
$t_1'(=t_1-\tau$) $\rightarrow$ $t_N'(=t_N-\tau)$ with the transit centred at 
time $t'=0$. The true anomaly at the time of transit minimum can be found by 
solving d$S$/d$f=0$ for $f$. For even a planet-only case, this is non-trivial 
and leads to a bi-quartic equation \citep{kipping:2008}. The solution may be 
found by a series expansion about the point of inferior conjunction 
$f(t=\tau) = \pi/2 - \omega + \eta$, where $\eta$ represents a perturbation 
term, which is given up to $6^{\mathrm{th}}$-order in \citet{thesis:2011}.

The third and final step is the application of Kepler's Equation. With the
reference true anomaly known, this is converted to a reference mean anomaly
via the usual relations \citep{murray:1999}. Then, the mean anomalies
at all times may be calculated since this parameter scales linearly with time.
Finally, the mean anomalies are converted into true anomalies using Kepler's
Equation.

So to summarize we have three steps: i) subtract the time of transit minimum 
from all times ii) define the reference anomaly as the time of transit minimum 
iii) assign mean anomalies for all times, which are then converted to true 
anomalies using Kepler's Equation. 

For data spanning multiple transits, the reference time should be close to the
weighted mean transit number. This selection typically minimizes the correlation
between the orbital period and reference time. In such a case, the range of
allowed values for the fitting routine to explore vary from 
$\tau = (\tau_{\mathrm{guess}} - P/2) \rightarrow (\tau_{\mathrm{guess}}+P/2)$.
Moving outside of this range would cause the fitting routine to assign a 
different transit epoch as the reference transit instead.

\subsubsection{The Planet-Moon Case}

Now consider a planet with a moon. The barycentre of the planet-moon system
essentially behaves in the same way which was used to describe the planet-only
case. The observed time of transit minimum of the planet is displaced from 
a linear ephemeris by a small time $\delta t$, due to transit timing
variations \citep{kipping:2009a}. However, the barycentre behaves identically to 
before and still passes the star at a minimum when 
$f_{B*}=(\pi/2-\omega_{B*}+\eta)$. Therefore, there is no need to change 
anything here except that it is understood that the $\tau$ value we fit for is 
not the time of transit minimum of the planet but the time of transit minimum of 
the planet-moon barycentre across the star, and thus we denote it as 
$\tau_{B*}$.

For the moon, an analogous logic may be followed. There must exist a second time 
shift, $\tau_{SB}$, to account for the phasing between the moon in its orbit
around the barycentre. It also seems clear that this phase time has a range of 
$P_{SB}$ (orbital period of the moon around the barycentre), in analogy to the 
$\tau_{B*}$ case. Let us imagine we have subtracted $\tau_{B*}$ from all of our 
times so are left with times running from, say, $t_1'\rightarrow t_N'$. The 
obvious deduction is that we must now subtract a value $\tau_{SB}$ to get to a 
new time frame $t''$, just as we did for $\tau_{B*}$.

A natural choice for $\tau_{SB}$ is the instant when d$S_{SB}$/d$f_{SB}=0$.
In analogy to the previous case, this occurs at 
$f_{SB} = \pi/2 - \omega_{SB} + \eta_{SB}$, where $\eta_{SB}$ is the
perturbation term. Because this parameter is a phasing term, we 
prefer to use $\phi_{SB} = (2 \pi \tau_{SB})/P_{SB}$ as a definition, to 
decrease the correlations with the orbital period.

\subsection{Retrograde Orbits}
\label{sub:retrograde}

A unique problem with \luna\ is that exomoons can be either prograde or 
retrograde. Whilst for planets this is also true, it actually makes no 
difference to the transit light curve (although the Rossiter-McLaughlin 
phenomenon is affected). However, for a moon the sense of motion is 
distinguishable via the TTV and TDV effects \citep{kipping:2009b}, meaning it 
does affect the transits and so cannot be neglected.

We tackle this by treating retrograde moons as having $\pi<i_{SB}<2\pi$ and
prograde moons as having $0<i_{SB}<\pi$. It is shown in \citet{thesis:2011}
that this definition produces the correct retrograde behaviour for the selected
coordinate system\footnote{Note that the longitude of the ascending node of
the moon in its orbit around the planet-moon barycentre has the range 
$0\leq\Omega_{SB}<2\pi$}.

We point out that the asymmetry between prograde and retrograde moons is small.
The source of the asymmetry comes from the relative phase difference between the
TDV-V and TDV-TIP effects, which is $0$ for prograde moons and $\pi$ for
retrograde \citep{kipping:2009b}. Therefore, the determination of the sense of 
orbital motion is only generally possible if both TDV-V and TDV-TIP effects are 
observable.

\subsection{Small-Moon Approximation}

From a naive perspective, one might start by considering that the transit light 
curve of a planet and a moon can be computed by calculating the light curve of 
both separately and then simply adding the signals together. However, there 
exists numerous scenarios where this would break down. For example, the moon 
could be eclipsing the star but fully inside the planetary disc and thus the 
change in flux caused by the moon would be zero.

To overcome this, we start by generating the planetary light curve in the normal 
way. This can be done using the \citet{mandel:2002} routine with any limb 
darkening law we wish and not making any size approximations. Having computed 
this curve, we need to add on the contribution from the moon. What really 
matters is what part of the moon is ``actively'' transiting the star. We define 
this as the area of the moon which overlaps the star but does not 
overlap the planet. If we can find this area, then we could compute a light 
curve for a planet + moon without any limb darkening effects immediately, 
which is a necessary first step. The subject of the moon's actively transiting 
area will be derived later in \S\ref{sec:areas}.

For now, let us assume we know what this area is and continue to think about how 
the limb darkened light curve of the actively transiting moon can be computed. 
In general, we are interested in achieving two things i) focussing on moons 
rather than, say, binary-planets\footnote{Although we intend to extend \luna\ 
to binary-planets in future-work.} ii) actually detecting such signals. The 
first of these means we are dealing with small objects such that $s\lesssim0.1$ 
in all cases and most likely $s\lesssim 0.01$ in most cases (where $s$ is 
the ratio of the satellite-to-star radii). The second point means that we need 
to be able to perform fits of transit light curves including all of the planet 
and moon properties. The inevitably large amount of parameter space necessitates 
computationally efficient and expedient algorithms. Putting these two arguments 
together indicates that the best way forward is to employ the small-planet 
approximation case used in \citet{mandel:2002}. This will ensure accurate 
modelling of the limb darkening but very fast algorithms. 

We stress that the planet is still modelled using the full equations since
the host planet could be a Jupiter sized object, yielding $p>0.1$ (where $p$
is the ratio of the planet-to-star radii). Therefore, we assume the moon is 
small and the planet is not. One limitation of this assumption is 
``binary-Jupiters''. If we have a binary-planet system for which the smaller 
body satisfies $s>0.1$ then the algorithm we develop here will be limited in 
accuracy. However, the current focus of \luna\ is solely to detect small 
bodies.


\subsection{Limb Darkened Light Curve for the Actively Transiting Lunar 
Component}
\subsubsection{Uniform source}

Let us assume that the component of the moon's area which actively transits the 
star is known and equals $A_{S,\mathrm{transit}}$. Before we can compute the 
resulting limb darkened light curve from this component, we must first consider 
the case for a uniform source.

We start by assuming the actively transiting component of the moon is equal
to that which would transit in the absence of a planet. In other words, we here
ignore the planet and will re-introduce it later. We follow the methodology of 
\citet{mandel:2002}, where the ratio of obscured to unobscured flux is given as 
$F_S^e(s,S_{S*}) = 1-\lambda_S^e(s,S_{S*})$ (replacing the appropriate symbols 
for the exomoon case):

\begin{equation}
\label{eqn:uniform}
\lambda_S^e(s,S_{S*}) = 
\left\{\begin{array}{ll}
0  &  1+s < S_{S*} \\
\frac{1}{\pi} \left[s^2 \kappa_0+\kappa_1-\kappa_2\right] &  |1-s| < S_{S*} 
\le 1+s \\
 s^2 & S_{S*} \le 1-s\\
1  &  S_{S*} \le s-1,\\
\end{array}\right.
\end{equation}
where $\kappa_1=\cos^{-1}[(1-s^2+S_{S*}^2)/2S_{S*}]$, 
$\kappa_0=\cos^{-1}[(s^2+S_{S*}^2-1)/2sS_{S*}]$ and 
$\kappa_2 = 0.5 \sqrt{4S_{S*}^2-(1+S_{S*}^2-s^2)^2}$.

The most interesting case is clearly $(1-s) \le S_{S*} \le (1+s)$, where only a 
portion of the moon transits the star. The area of the moon which transits the 
star may be denoted as $\alpha_{S*}$ where the subscript indicates $S$ is 
transiting $*$.

In this case, $\lambda_S^e = \alpha_{S*}/\pi$, but $\alpha$ really generally 
describes the area of intersection between any two circles. We prefer to 
generalize the equations at this stage, which will prove useful later. 
Therefore, the area of intersection caused by object of radius $r$ transiting 
object of radius $R$, with separation $S$, is:

\begin{align}
\alpha(R,r,S) &= r^2 \kappa_0(R,r,S) + R^2 \kappa_1(R,r,S) - \kappa_2(R,r,S) \\
\kappa_0(R,r,S) &= \arccos\Big[\frac{S^2+r^2-R^2}{2 S r}\Big] \\
\kappa_1(R,r,S) &= \arccos\Big[\frac{S^2+R^2-r^2}{2 S R}\Big] \\
\kappa_2(R,r,S) &= \sqrt{\frac{4 S^2 R^2 - (R^2+S^2-r^2)^2}{4}}
\label{eqn:alpha}
\end{align}

Let us now re-introduce the planet. The consequences are that in various
circumstances the portion of the moon's shadow which actively transits the star
is diminished due to overlap between the planet and the moon. We will discuss
the derivation of the actively transiting lunar area in \S\ref{sec:areas}, but
for now it is sufficient to say that for a uniform source the loss of light
due to the lunar component is:

\begin{equation}
F_S^e(s,S_{S*}) = 1-\frac{ A_{S,\mathrm{transit}} }{\pi}
\end{equation}

As an example, in the case of no overlap between the planet and the moon, but
the moon fully inside the stellar disc, $A_{S,\mathrm{transit}}=\pi s^2$ and 
thus we recover the expected form shown in Equation~\ref{eqn:uniform}.

\subsubsection{List of Cases}

Looking at the equations for a uniform source (Equation~\ref{eqn:uniform}), one 
may identify three distinct cases: i) moon fully outside the stellar disc 
ii) moon partially inside the stellar disc iii) moon fully inside the 
stellar disc. The first of these is trivial; there is no transit occurring. The 
third is also trivial for a uniform source but not so for a limb darkened 
one. For reasons which become clearer later, it is necessary to separate 
ii) into two different cases dependent upon whether the moon overlaps the 
stellar centre or not (cases III and IX in the original \citet{mandel:2002}
assignment). To summarize the different cases of interest, we direct 
the reader to Table~\ref{tab:mandel}.

\begin{table}									%
\caption{List of cases identified by \citet{mandel:2002}. We use the same case	%
notation in this work, but modifying them to emphasise the focus on moons.}	%
\centering 
\begin{tabular}{l c c} 
\hline\hline 
Case & Condition & Area of star which is eclipsed \\ [0.5ex] 			%
\hline 
I & $1+s<S_{S*}<\infty$ & $0$ \\						%
II & $1-s<S_{S*}<1+s$ & $\alpha_{S*}$ \\					%
III & $s<S_{S*}<1-s$ & $\pi s^2$ \\						%
IX & $0<S_{S*}<s$ & $\pi s^2$ \\ [1ex]						%
\hline\hline 
\end{tabular}									%
\label{tab:mandel} 
\end{table}									%

\subsubsection{Case III}

Under the assumption of a small ratio-of-radii ($s \ll 1$) one may assume the 
surface brightness of the star is constant under the disc of the transiting body 
\citep{mandel:2002}. We will here assume non-linear limb darkening of the form
$I(r) = 1 - \sum_{n=1}^4 c_n (1-\mu^{n/2})$, where $I(r)$ represents the radial 
intensity distribution from the star with $I(0) = 1$, 
$\mu = \cos \theta = \sqrt{1-r^2}$ and $r$ is the normalized radial coordinate 
on the disc of the star. The non-linear limb darkening law \citep{claret:2000}
is chosen as it may be easily used to compute lower order limb darkening laws,
specifically quadratic and linear laws.

Let us draw an annulus inside the star with width $2s$ at 
radius $S_{S*}$ to represent this surface. We begin by considering case III, 
where the body is fully inside the star but does not cover the stellar centre 
(i.e. $s<S_{S*}<1-s$). We have:

\begin{align}
F_{\mathrm{total}} &= \int_{0}^{1} 2 r I(r) \, \mathrm{d}r \nonumber \\
\qquad&= 1 - \sum_{n=1}^4 \frac{n c_n}{n+4} \\
F_{S,\mathrm{annulus}}^{\mathrm{III}}(S_{S*},s) &= \int_{S_{S*}-s}^{S_{S*}+s} 
2 r I(r) \, \mathrm{d}r \nonumber \\
\qquad&= \frac{4}{5} c_1 \Bigg[s^2 \left(\sqrt[4]{1-b_m}-\sqrt[4]{1-a_m}\right) 
\nonumber \\
\qquad& +\left(S_{S*}^2-1\right) \left(\sqrt[4]{1-b_m}-\sqrt[4]{1-a_m }\right) 
\nonumber \\
\qquad& +s S_{S*} \left(2\sqrt[4]{1-a_m}+2 \sqrt[4]{1-b_m}-5\right)\Bigg] 
\nonumber \\
\qquad& +\frac{2}{21} \Bigg[7 c_2 \Big[(s^2 \left(\sqrt{1-b_m}-
\sqrt{1-a_m}\right) \nonumber \\
\qquad& +\left(S_{S*}^2-1\right) \left(\sqrt{1-b_m}-\sqrt{1-a_m}\right) 
\nonumber \\
\qquad& +2 s S_{S*} \left(\sqrt{1-a_m}+\sqrt{1-b_m}-3\right)\Big] \nonumber \\
\qquad& -6 \Big[c_3 \Big(s^2 \left(\left(1-a_m\right)^{3/4}-(1-b_m)^{3/4}\right) 
\nonumber \\
\qquad& +S_{S*}^2\left(\left(1-a_m\right)^{3/4}-(1-b_m)^{3/4}\right) 
\nonumber \\
\qquad& +s S_{S*} \left(-2 \left(1-a_m\right)^{3/4}-2 (1-b_m)^{3/4}+7\right) 
\nonumber \\
\qquad& -\left(1-a_m\right)^{3/4}+(1-b_m)^{3/4}\Big) \nonumber \\
\qquad& +7 s S_{S*} \left(c_4\left(s^2+S_{S*}^2\right)-1\right)\Big]\Bigg]
\end{align}

where $a_m = (S_{S*}-s)^2$ and $b_m = (S_{S*}+s)^2$ 
and the $m$ subscript denotes this definition comes from \citet{mandel:2002}. 
A further simplification is possible by using $a_{mr} = (1-a_m)^{1/4}$ and 
$b_{mr} = (1-b_m)^{1/4}$.

\begin{align}
F_{S,\mathrm{annulus}}^{\mathrm{III}}(S_{S*},s) &= \frac{4}{5} \Big[ s^2 
(b_{mr}-a_{mr}) + (S_{S*}^2-1) (b_{mr}-a_{mr}) \nonumber \\
\qquad& + 2 s S_{S*} \Big( -\frac{5}{2} + a_{mr} + b_{mr} \Big) \Big] c_1 
\nonumber \\
\qquad& + \frac{2}{21} \Bigg[ 7 \Big( s^2 (b_{mr}^2-a_{mr}^2) + (S_{S*}^2-1) 
(b_{mr}^2-a_{mr}^2) \nonumber \\
\qquad& + 2 s S_{S*} (-3+a_{mr}^2 + b_{mr}^2) \Big) c_2 \nonumber \\
\qquad& -6 \Bigg( \Big( b_{mr}^3-a_{mr}^3 + 2 s S_{S*} (\frac{7}{2} - a_{mr}^3 - 
b_{mr}^3) \nonumber \\
\qquad& + s^2 (a_{mr}^3 - b_{mr}^3) + S_{S*}^2 (a_{mr}^3 - b_{mr}^3) \Big) c_3 
\nonumber \\
\qquad& + 7 s S_{S*} (-1+(s^2+S_{S*}^2) c_4) \Bigg) \Bigg]
\end{align}

The final step is to acknowledge that the transiting body does not transit an 
area equal to that of the annulus, but in general, a smaller area. Therefore, we
must multiply the flux from the annulus by a ratio-of-areas given by:

\begin{align}
\mathcal{A}_{S}^{\mathrm{III}} &= 
\frac{A_{S,\mathrm{transit}}}{A_{S,\mathrm{annulus}}^{\mathrm{III}}} \\ 
\qquad&= \frac{ A_{S,\mathrm{transit}} }{\pi [(S_{S*}+s)^2 - (S_{S*}-s)^2]} 
\nonumber \\
\mathcal{A}_{S}^{\mathrm{III}} &= \frac{ A_{S,\mathrm{transit}} }{\pi (b_m-a_m)}
\end{align}

The final flux is given by:

\begin{equation}
F_{S}^{e} = 1 - \mathcal{A}_{S}^{e} 
\Big(\frac{ F_{S,\mathrm{annulus}}^{e} }{F_{\mathrm{total}}}\Big)
\label{eqn:fluxeqne}
\end{equation}

\subsubsection{Case IX}

If the transiting body has an equatorial orbit, such that $S_{S*}<s$, the setup 
for case III is invalid since one of the integration limits is negative and $r$ 
is strictly defined to be $r\geq0$. This gives us case IX:

\begin{align}
F_{S,\mathrm{annulus}}^{\mathrm{IX}} &= \int_{0}^{S_{S*}+s} 2 r I(r) \, 
\mathrm{d}r \nonumber \\
\qquad&= \frac{4}{5} \Bigg[ 1 - b_{mr} + (s+S_{S*})^2 \Big( 
-\frac{5}{4}+b_{mr}\Big) \Bigg] c_1 \nonumber \\
\qquad& + \frac{2}{3} \Bigg[ 1 - b_{mr}^2 + (s+S_{S*})^2 \Big(-\frac{3}{2} + 
b_{mr}^2\Big) \Bigg] c_2 \nonumber \\
\qquad& + \frac{1}{14} \Bigg[14 (s+S_{S*})^2 - 7 (s+S_{S*})^4 c_4 \nonumber \\
\qquad& + 2 \Big(4 - 4 b_{mr}^3 + (s+S_{S*})^2 (-7+4 b_{mr}^3)\Big) c_3 \Bigg]
\end{align}

For the ratio of the annulus to actively-transiting-lunar area we have:

\begin{align}
\mathcal{A}^{\mathrm{IX}} &= \frac{A_{S,\mathrm{transit}}}{\pi b_m}
\end{align}

\subsubsection{Case II}

The final case we need is case II. This is when the body is in the 
ingress/egress portion i.e. $(1-s)<S_{S*}<(1+s)$. In this case the annulus flux 
is given by:

\begin{align}
F_{S,\mathrm{annulus}}^{\mathrm{II}} &= \int_{S_{S*}-s}^{1} 2 r I(r) \, 
\mathrm{d}r \nonumber \\
\qquad&= \frac{1}{5} ( -5 + 4 a_{mr} ) a_{mr}^4 c_1 \nonumber \\
\qquad& + \frac{a_m-1}{42} \Bigg[ -42 + (42-28 a_{mr}^2) c_2 + 6(7-4 a_{mr}^3) c_3 
\nonumber \\
\qquad& + 21 (1+a_m) c_4 \Bigg]
\end{align}

The ratio of the annulus to actively-transiting-lunar area is:

\begin{align}
\mathcal{A}^{\mathrm{II}} &= \frac{ A_{S,\mathrm{transit}} }{\pi (1-a_m)}
\end{align}

\subsection{Summary}

The light curve of the actively-transiting-lunar component has now been
evaluated, but we have not yet derived $A_{S,\mathrm{transit}}$: the 
sky-projected area of this component. The planetary light curve is computed
using the usual \citet{mandel:2002} routine, utilizing $S_{P*}$. The only
remaining task is therefore to find $A_{S,\mathrm{transit}}$, which we deal
with next in \S\ref{sec:areas}.



\section{Actively Transiting Lunar Area}
\label{sec:areas}

\subsection{Principal Cases}

The one-body case is described by one parameter, $S$, which can have three 
states. As we saw in \S\ref{sec:lightcurves}, the light curve of the 
actively-transiting-lunar component can be generated in each case by just 
knowing $S_{S*}$ and $A_{S,\mathrm{transit}}$, in the small moon approximation. 
In what follows, we always assume $s^2<p^2<1$ where $s = R_S/R_*$. It is 
therefore clear that our task is to find $A_{S,\mathrm{transit}}$ in all 
possible configurations i.e. the actively transiting area. This will enable us 
to compute $\mathcal{A}^e$ in cases II, III and IX and thus produce limb 
darkened transit light curves for the moon.

\begin{figure*}
\begin{center}
\includegraphics[width=16.8 cm]{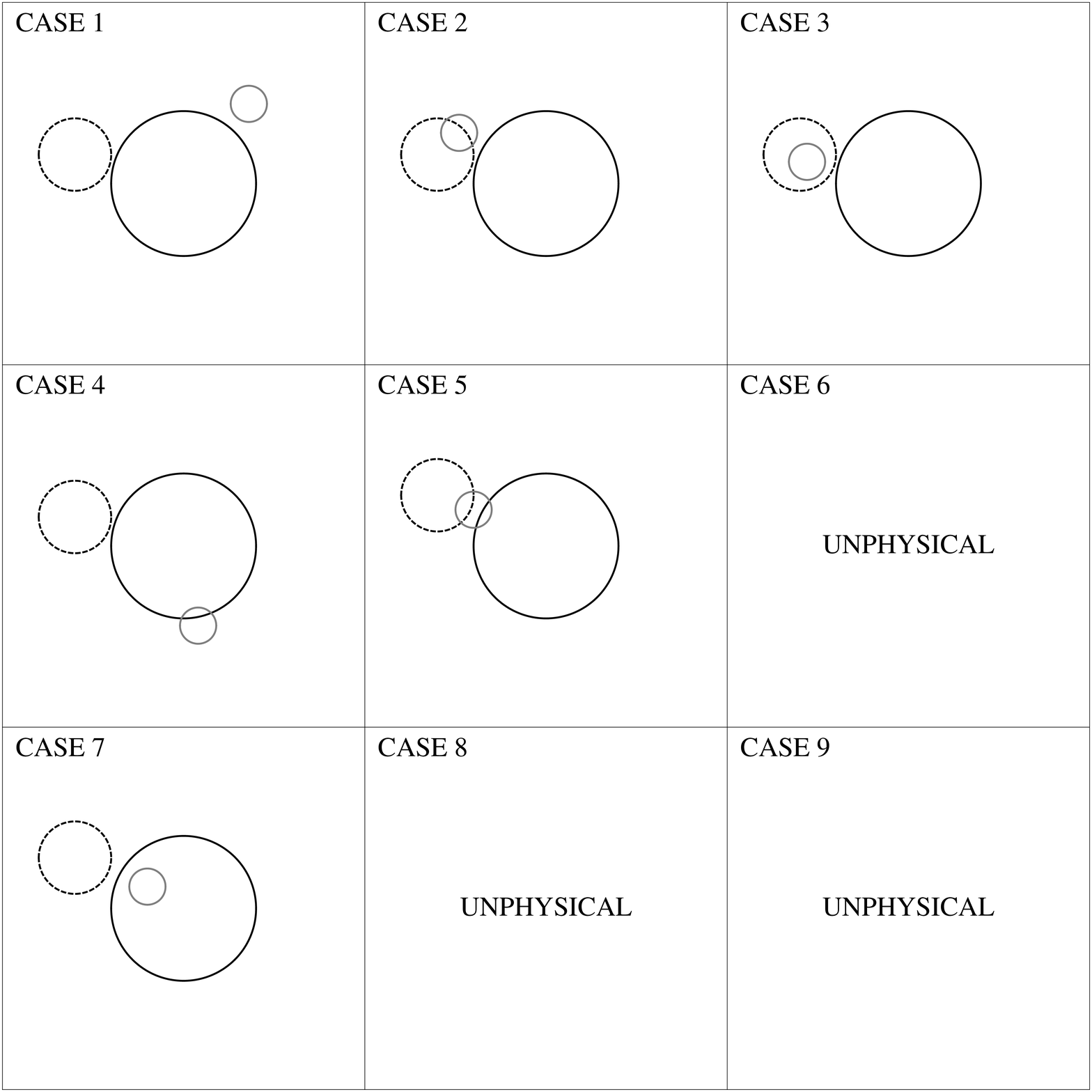}
\caption{\emph{Principal cases 1 to 9. The star is black solid,
the planet is black dashed and the moon is gray solid.}} 
\label{fig:prin1to9}
\end{center}
\end{figure*}

\begin{figure*}
\begin{center}
\includegraphics[width=16.8 cm]{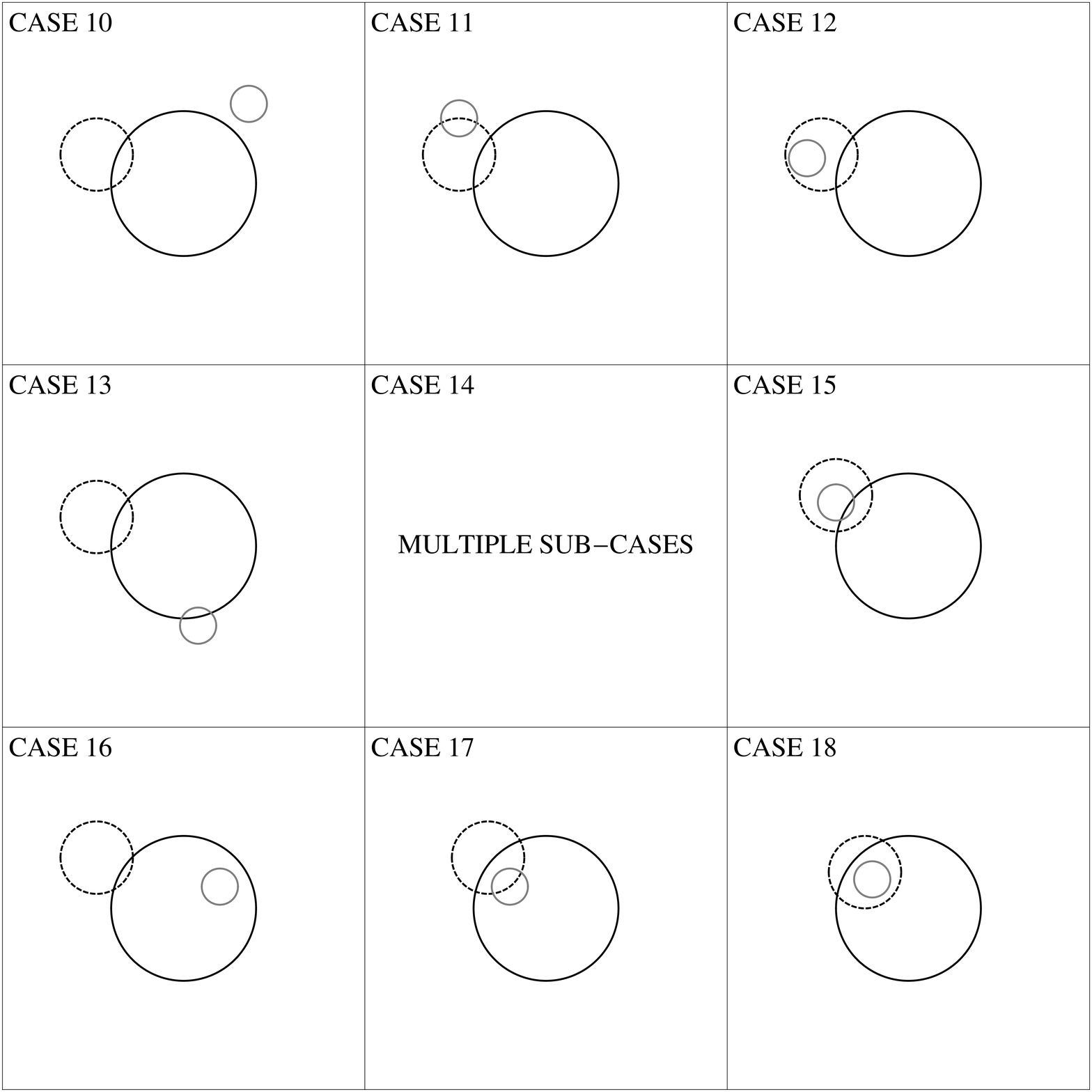}
\caption{\emph{Principal cases 10 to 18. The star is black solid,
the planet is black dashed and the moon is gray solid.}} 
\label{fig:prin10to18}
\end{center}
\end{figure*}

\begin{figure*}
\begin{center}
\includegraphics[width=16.8 cm]{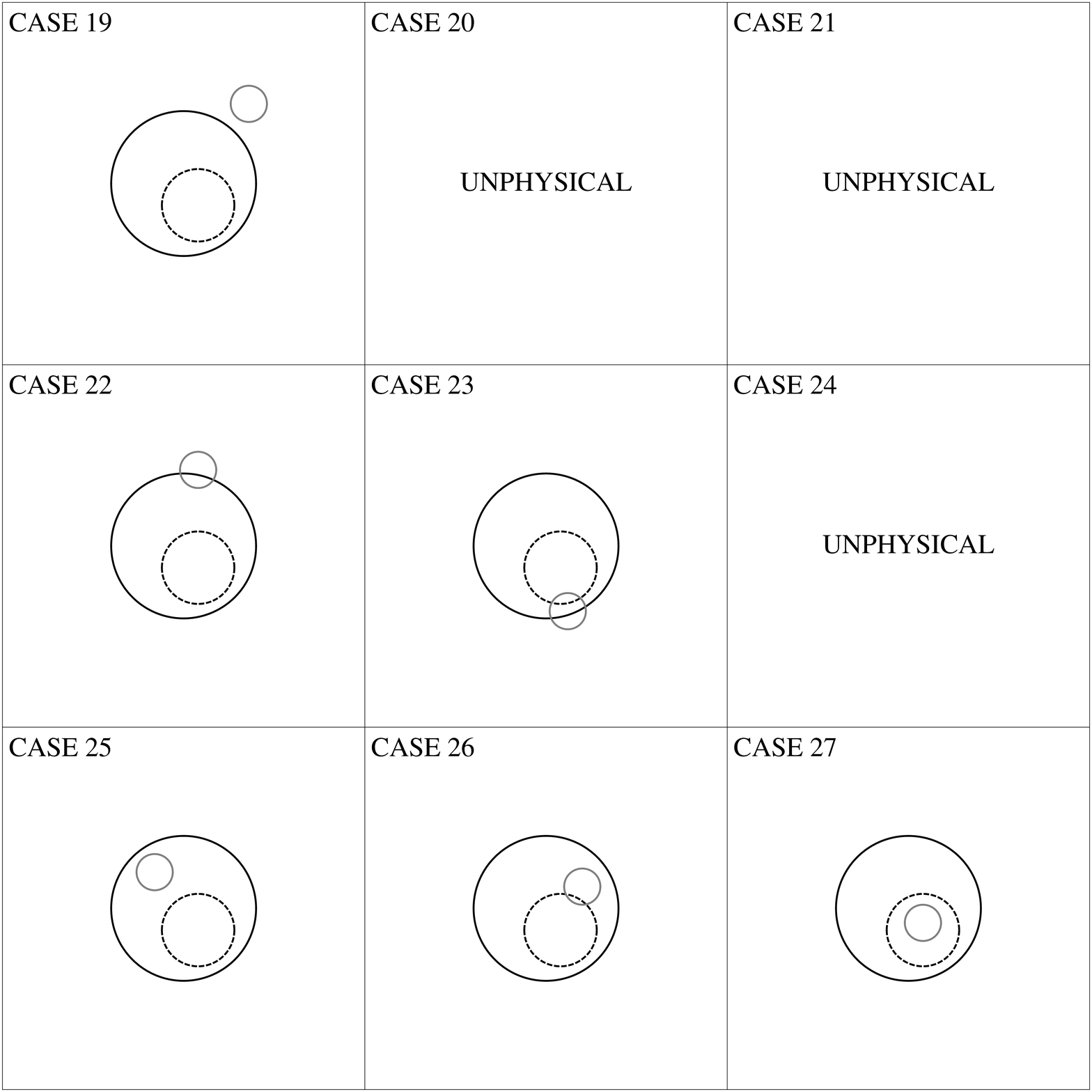}
\caption{\emph{Principal cases 19 to 27. The star is black solid,
the planet is black dashed and the moon is gray solid.}} 
\label{fig:prin19to27}
\end{center}
\end{figure*}

With three $S$ values now in play ($S_{P*}$, $S_{S*}$ \& $S_{SP}$) more cases 
are possible than before. In total $3^3 = 27$ principal cases exist, but many of 
these are unphysical under the conditions that $R>p>s$ (where $R=1$) and that 
$S_{PS} < |S_{P*} - S_{S*}|$. In Table~\ref{tab:principals}, we list all the 
possible cases by permuting each $S$ parameter into each of their respective 
three states. The cases which are unphysical are marked accordingly. 
Figures~\ref{fig:prin1to9}, \ref{fig:prin10to18} \& \ref{fig:prin19to27} 
illustrate example configurations for each principal case.

Table~\ref{tab:principals} shows that for some cases, the amount of lunar area 
actively transiting the star is dependent upon the position of the planet (these 
case are marked with a *). This is because the planet's shadow overlaps with the 
lunar shadow and so the the moon cannot block out to its full potential. We 
label these cases as exhibiting ``areal interaction''.

\begin{table*}
\caption{List of principal cases. For each case we label whether the 
case is geometrically physical or not. We also give the expected eclipsed area 
due to each body. More complicated cases are marked with a * and 
exhibit areal interaction. ** indicates multiple sub-cases exist, which is 
discussed in \ref{sub:case14subcases}.} 
\centering 
\begin{tabular}{c c c c c c c} 
\hline\hline 
Case Number [$\mathcal{E}$] & $S_P$ & $S_S$ & $S_{PS}$ & Physical? & 
$A_{P,\mathrm{transit}}$ & $A_{S,\mathrm{transit}}$ \\ [0.5ex] 
\hline 
1 & $S_{P*} \ge 1+p$ & $S_{S*} \ge 1+s$ & $S_{PS} \ge p+s$ & $\surd$ & $0$ & 
$0$ \\
2 & $S_{P*} \ge 1+p$ & $S_{S*} \ge 1+s$ & $p-s < S_{PS} < p+s$ & $\surd$ & $0$ 
& $0$ \\
3 & $S_{P*} \ge 1+p$ & $S_{S*} \ge 1+s$ & $S_{PS} \le p-s$ & $\surd$ & $0$ & 
$0$ \\
4 & $S_{P*} \ge 1+p$ & $1-s < S_{S*} < 1+s$ & $S_{PS} \ge p+s$ & $\surd$ & $0$ 
& $\alpha_{S*}$ \\ 
5 & $S_{P*} \ge 1+p$ & $1-s < S_{S*} < 1+s$ & $p-s < S_{PS} < p+s$ & $\surd$ & 
$0$ & $\alpha_{S*}$ \\
6 & $S_{P*} \ge 1+p$ & $1-s < S_{S*} < 1+s$ & $S_{PS} \le p-s$ & $\times$ & - 
& - \\
7 & $S_{P*} \ge 1+p$ & $S_{S*} \le 1-s$ & $S_{PS} \ge p+s$ & $\surd$ & $0$ & 
$\pi s^2$ \\
8 & $S_{P*} \ge 1+p$ & $S_{S*} \le 1-s$ & $p-s < S_{PS} < p+s$ & $\times$ & - 
& - \\
9 & $S_{P*} \ge 1+p$ & $S_{S*} \le 1-s$ & $S_{PS} \le p-s$ & $\times$ & - & - \\
\hline
10 & $1-p < S_{P*} < 1+p$ & $S_{S*} \ge 1+s$ & $S_{PS} \ge p+s$ & $\surd$ & 
$\alpha_{P*}$ & $0$ \\
11 & $1-p < S_{P*} < 1+p$ & $S_{S*} \ge 1+s$ & $p-s < S_{PS} < p+s$ & $\surd$ & 
$\alpha_{P*}$  & $0$ \\
12 & $1-p < S_{P*} < 1+p$ & $S_{S*} \ge 1+s$ & $S_{PS} \le p-s$ & $\surd$ & 
$\alpha_{P*}$  & $0$ \\
13 & $1-p < S_{P*} < 1+p$ & $1-s < S_{S*} < 1+s$ & $S_{PS} \ge p+s$ & $\surd$ & 
$\alpha_{P*}$  & $\alpha_{S*}$ \\ 
14** & $1-p < S_{P*} < 1+p$ & $1-s < S_{S*} < 1+s$ & $p-s < S_{PS} < p+s$ & 
$\surd$ & $\alpha_{P*}$  & variable \\
15 & $1-p < S_{P*} < 1+p$ & $1-s < S_{S*} < 1+s$ & $S_{PS} \le p-s$ & $\surd$ & 
$\alpha_{P*}$  & $0$ \\
16 & $1-p < S_{P*} < 1+p$ & $S_{S*} \le 1-s$ & $S_{PS} \ge p+s$ & $\surd$ & 
$\alpha_{P*}$  & $\pi s^2$ \\
17* & $1-p < S_{P*} < 1+p$ & $S_{S*} \le 1-s$ & $p-s < S_{PS} < p+s$ & $\surd$ 
& $\alpha_{P*}$ & $\pi s^2 - \alpha_{SP}$ \\
18 & $1-p < S_{P*} < 1+p$ & $S_{S*} \le 1-s$ & $S_{PS} \le p-s$ & $\surd$ & 
$\alpha_{P*}$ & 0 \\
\hline
19 & $S_{P*} \le 1-p$ & $S_{S*} \ge 1+s$ & $S_{PS} \ge p+s$ & $\surd$ & 
$\pi p^2$ & $0$ \\
20 & $S_{P*} \le 1-p$ & $S_{S*} \ge 1+s$ & $p-s < S_{PS} < p+s$ & $\times$ & - 
& - \\
21 & $S_{P*} \le 1-p$ & $S_{S*} \ge 1+s$ & $S_{PS} \le p-s$ & $\times$ & - 
& - \\
22 & $S_{P*} \le 1-p$ & $1-s < S_{S*} < 1+s$ & $S_{PS} \ge p+s$ & $\surd$ & 
$\pi p^2$ & $\alpha_{S*}$ \\ 
23* & $S_{P*} \le 1-p$ & $1-s < S_{S*} < 1+s$ & $p-s < S_{PS} < p+s$ & $\surd$ 
& $\pi p^2$ & $\alpha_{S*} - \alpha_{SP}$ \\
24 & $S_{P*} \le 1-p$ & $1-s < S_{S*} < 1+s$ & $S_{PS} \le p-s$ & $\times$ & - 
& - \\
25 & $S_{P*} \le 1-p$ & $S_{S*} \le 1-s$ & $S_{PS} \ge p+s$ & $\surd$ & 
$\pi p^2$ & $\pi s^2$ \\
26* & $S_{P*} \le 1-p$ & $S_{S*} \le 1-s$ & $p-s < S_{PS} < p+s$ & $\surd$ & 
$\pi p^2$ & $\pi s^2 - \alpha_{SP}$ \\
27 & $S_{P*} \le 1-p$ & $S_{S*} \le 1-s$ & $S_{PS} \le p-s$ & $\surd$ & 
$\pi p^2$ & $0$ \\ [1ex]
\hline\hline 
\end{tabular}
\label{tab:principals} 
\end{table*}


\subsubsection{Case 26}

We start by considering the simplest of the areal interaction cases, case 26 
(see Figure~\ref{fig:prin19to27}). We denote this as $\mathcal{E}26$ in 
mathematical notation (and likewise for the other case numbers). $\mathcal{E}$ 
is chosen in analogy to the case notation $e$ used by \citet{mandel:2002}. For 
case 26, the planet and moon are both completely inside the stellar disc but are 
interacting with each other. This is very similar to the case II situation for a 
one-body transit where the star has become the planet and the transiting-body 
has become the moon. So $\{R,p,S\} \rightarrow \{p,s,S_{PS}\}$. The area of the 
moon which left actively transiting the star is given by 
$A_{S,\mathrm{transit}}^{\mathcal{E}26} = \pi s^2 - \alpha_{SP}$.

\subsubsection{Case 17}

The next most complicated case is case 17. Here the planet is on the limb of the 
star and the moon is on the limb of the planet, but completely within the star. 
It can be seen that 
$A_{S,\mathrm{transit}}^{\mathcal{E}27} = A_{S,\mathrm{transit}}^{\mathcal{E}26} 
= \pi s^2 - \alpha_{SP}$ 
and thus we already have the required equation in hand.

\subsubsection{Case 23}

The third most complicated case is when the planet is completely inside the star 
but the moon sits on the planetary limb and coincidentally the stellar limb. If 
the planet was not present, the moon would eclipse an area given by 
$\alpha_{S*}$.

However, the planet is present, and it blocks out a portion of this area. Under 
the conditions of case 23, the overlapping area between the planet and moon must 
always lie completely within the stellar disc since the planet always lies 
completely within the stellar disc. It can be therefore be seen that the 
overlapping area, $\alpha_{SP}$, will always be less than $\alpha_{S*}$. 
Therefore, the final area which is actively transiting the star from the moon 
alone is:

\begin{align}
A_{S,\mathrm{transit}}^{\mathcal{E}23} &= \alpha_{S*} - \alpha_{SP}
\end{align}

\subsection{The Sub-Cases of Case 14}
\label{sub:case14subcases}

Case 14 is the most complicated case to consider. Indeed, case 14 actually has 
multiple sub-cases which define the various possible behaviours. Although 
similar to case 23, the planet's shadow now does not completely eclipse the star 
and so the overlapping area between the moon and planet also does not 
necessarily have to lie completely within the stellar disc. As a consequence, 
multiple different states exist and a singular expression is not possible for 
$A_{S,\mathrm{transit}}$.

The first key question is: \emph{how many sub-cases actually exist}? The problem 
we are thinking of is that of three circles interacting to give an overlapping 
region, for which we require the associated area. The problem of 
finding the area of overlap between three circles was studied extensively in the
pioneering work of \citet{fewell:2006}. Indeed, an analytic solution for the 
overlap of three circles did not exist until 2006 when Fewell presented the 
solution. \citet{fewell:2006} identified nine possible cases in total, which we 
label as Fewell case (1) $\rightarrow$ (9) and viewable in Figure~5 of 
\citet{fewell:2006}. To assign identities in our framework, we label the biggest 
circle the star, the middle-sized circle the planet and the smallest circle the 
moon in Figure~5 of \citet{fewell:2006}.

Going through the \citet{fewell:2006} scenarios with our assigned identities, 
the first thing we notice is that only four of the \citet{fewell:2006} cases 
remain consistent with the case 14 conditions, the others are much simpler 
scenarios which we dealt with earlier as the principal cases. In 
Table~\ref{tab:fewell}, we show the corresponding case numbers to the 
\citet{fewell:2006} cases for completion.

\begin{table}
\caption{List of \citet{fewell:2006} cases and the corresponding case 
definitions used in this work.} 
\centering 
\begin{tabular}{c c} 
\hline\hline 
\citet{fewell:2006} Case Number & Case Number (This work) 
[$\mathcal{E}$] \\ [0.5ex] 
\hline 
(1) & 14.1 \\
(2) & 14.2 \\
(3) & 14.3 \\
(4) & 17 \\
(5) & 18 \\
(6) & 27 \\
(7) & 14.7 \\
(8) & 11 \\
(9) & 10 \\ [1ex]
\hline\hline 
\end{tabular}
\label{tab:fewell} 
\end{table}

The \citet{fewell:2006} solution for the area of common overlap between three
circles is based upon a formula for the circular triangle, which uses the chord 
lengths and radii of the three circles involved. Only in Fewell case (1) does a 
circular triangle actually exist and this is the focus of \citet{fewell:2006}. 
However, \citet{fewell:2006} performs the derivation by calculating analytic 
expressions for the vertices of all three circles which may then be converted to 
chord lengths and these equations are useful for distinguishing between all the 
various sub-cases.

Let us consider three circles labelled 1, 2 and 3 of radii $\mathcal{R}_i$ and 
positions $\{\mathcal{X}_i,\mathcal{Y}_i\}$ and intersection points 
$\{\mathcal{X}_{ij},\mathcal{Y}_{ij}\}$ (where two intersections points always 
exist for each $i$-$j$ pairing under the case 14 conditions). Let us define 
$\{\mathcal{X}_1,\mathcal{Y}_1\} = \{0,0\}$ and assume 
$\mathcal{R}_1 > \mathcal{R}_2 > \mathcal{R}_3$. Therefore, we have circle 
1 being the star, circle 2 being the planet and circle 3 being the moon. The 
intersection points are given by (see \citet{fewell:2006} for details of the
relevant derivation):

\begin{align}
&\mathbf{Intersection\,\,12} \nonumber \\
\mathcal{X}_{12} &= \frac{1 - p^2 + S_{P*}^2}{2 S_{P*}} \\
\mathcal{Y}_{12} &= \frac{1}{2 S_{P*}} \sqrt{2 S_{P*}^2 (1 + p^2) - (1 - p^2)^2 
- S_{P*}^4}
\end{align}

\begin{align}
&\mathbf{Intersection\,\,13} \nonumber \\
\mathcal{X}_{13}  &= \mathcal{X}_{13}'\cos\theta' - 
\mathcal{Y}_{13}'\sin\theta' \\
\mathcal{Y}_{13}  &= \mathcal{X}_{13}'\sin\theta' + 
\mathcal{Y}_{13}'\cos\theta' \\
\mathcal{X}_{13}' &= \frac{1 - s^2 + S_{S*}^2}{2 S_{S*}} \\
\mathcal{Y}_{13}' &= \frac{-1}{2S_{S*}} \sqrt{2 S_{S*}^2(1+s^2) - (1 -s^2)^2 - 
S_{S*}^4} \\
\cos\theta' &= \frac{S_{P*}^2 + S_{S*}^2 - S_{PS}^2}{2 S_{P*} S_{S*}} \\
\sin\theta' &= \sqrt{1-\cos^2\theta'}
\end{align}

\begin{align}
&\mathbf{Intersection\,\,23} \nonumber \\
\mathcal{X}_{23}       &= \mathcal{X}_{23}'' \cos\theta'' - 
\mathcal{Y}_{23}''\sin\theta'' + S_{P*} \\
\mathcal{Y}_{23}       &= \mathcal{X}_{23}'' \sin\theta'' + 
\mathcal{Y}_{23}''\cos\theta''          \\
\mathcal{X}_{23}''     &= \frac{p^2 - s^2 + S_{PS}^2}{2 S_{PS}} \\
\mathcal{Y}_{23}''     &= \frac{1}{2 S_{PS}} \sqrt{2 S_{PS}^2 (p^2 + s^2) - 
(p^2 - s^2)^2 - S_{PS}^4} \\
\cos\theta'' &= -\frac{S_{P*}^2 + S_{PS}^2 - S_{S*}^2}{2 S_{P*} S_{PS}} \\
\sin\theta'' &= \sqrt{1-\cos^2\theta''}
\end{align}



\subsubsection{Fewell Case (1)}

What makes Fewell case (1) unique mathematically speaking? i.e. under what 
conditions are the equations for (1) valid? \citet{fewell:2006} provides 
expressions for evaluating whether case (1) is satisfied or not. 
\citet{fewell:2006} identifies ``two'' conditions but one of these conditions is 
just the case 14 conditions and need not be considered again. The second 
condition is actually composed of two causal factors and so in total there 
exists two conditions which uniquely define case 14.1. Let us define conditions 
A and B as:

\begin{align}
&\mathbf{Condition\,\,A} \nonumber \\
&(\mathcal{X}_{12} - S_{S*} \cos \theta')^2 + (\mathcal{Y}_{12} - 
S_{S*} \sin \theta')^2 < s^2
\label{eqn:conditionA}
\end{align}

\begin{align}
&\mathbf{Condition\,\,B} \nonumber \\
&(\mathcal{X}_{12} - S_{S*} \cos \theta')^2 + (\mathcal{Y}_{12} 
+ S_{S*} \sin \theta')^2 < s^2
\label{eqn:conditionB}
\end{align}

In order to have case 14.1, we require that condition A is satisfied and 
condition B is anti-satisfied (we call this condition $\bar{B}$). The area 
$\alpha_{\mathrm{transit}}$ may be found by evaluating the circular triangle, 
principally defined by the chord lengths and radii. In fact, two 
sub-sub-cases exist here for 14.1 which we label as 14.1a and 14.1b. 14.1a is 
the case where less than half the circle is involved and 14.1b occurs when more 
than half the circle is involved. The two cases are distinguished by evaluating 
the following:

\begin{align}
&\mathbf{Condition\,\,C} \nonumber \\
&S_{S*} \sin \theta' > \mathcal{Y}_{13} + \frac{\mathcal{Y}_{23} - 
\mathcal{Y}_{13}}{\mathcal{X}_{23} - \mathcal{X}_{13}} (S_{S*} \cos \theta' - 
\mathcal{X}_{13})
\label{eqn:conditionC}
\end{align}

If the above equation is true (condition C) then we have case 14.1a and 
$A_{\mathrm{overlap}}^{\mathcal{E}14.1a}$ may be found using:

\begin{align}
A_{\mathrm{overlap}}^{\mathcal{E}14.1a} &= \frac{1}{4} \sqrt{(\mathfrak{c}_1 + 
\mathfrak{c}_2 + \mathfrak{c}_3) (\mathfrak{c}_2 + \mathfrak{c}_3 - \mathfrak{c}_1)} \nonumber\\
\qquad& \times \sqrt{(\mathfrak{c}_1 + \mathfrak{c}_3 - 
\mathfrak{c}_2) (\mathfrak{c}_1 + \mathfrak{c}_2 - \mathfrak{c}_3)} \nonumber \\
\qquad& + \sum_{k=1}^3 \Big(\mathcal{R}_k^2 
\arcsin \frac{\mathfrak{c}_k}{2 \mathcal{R}_k} - \frac{\mathfrak{c}_k}{4} 
\sqrt{4 \mathcal{R}_k^2 - \mathfrak{c}_k^2}\Big) \\
\mathfrak{c}_k^2 &= (\mathcal{X}_{ik} - \mathcal{X}_{jk})^2 + (\mathcal{Y}_{ik} 
- \mathcal{Y}_{jk})^2
\end{align}

Otherwise (condition $\bar{C}$), for case 14.1b, we must use:

\begin{align}
A_{\mathrm{overlap}}^{\mathcal{E}14.1b} &= \frac{1}{4} \sqrt{(\mathfrak{c}_1 + 
\mathfrak{c}_2 + \mathfrak{c}_3) (\mathfrak{c}_2 + \mathfrak{c}_3 - 
\mathfrak{c}_1)} \nonumber \\
\qquad& \times \sqrt{(\mathfrak{c}_1 + \mathfrak{c}_3 - 
\mathfrak{c}_2) (\mathfrak{c}_1 + \mathfrak{c}_2 - \mathfrak{c}_3)} \nonumber \\
\qquad& + \sum_{k=1}^3 \Big(\mathcal{R}_k^2 \arcsin \frac{\mathfrak{c}_k}{2 
\mathcal{R}_k}\Big) \nonumber \\
\qquad& - \frac{\mathfrak{c}_1}{4} \sqrt{4 \mathcal{R}_1^2 - \mathfrak{c}_1^2}- 
\frac{\mathfrak{c}_2}{4} \sqrt{4 \mathcal{R}_2^2 - \mathfrak{c}_2^2} + 
\frac{\mathfrak{c}_3}{4} \sqrt{4 \mathcal{R}_3^2 - \mathfrak{c}_3^2}
\end{align}

The actively transiting area of the moon is now given by: 
$A_{S,\mathrm{transit}}^{\mathcal{E}14.1x} = \alpha_{S*} - 
A_{\mathrm{overlap}}^{\mathcal{E}14.1x}$. 
In the above equations, it should be noted that $\mathfrak{c}_k$ represents the 
chord lengths.

\subsubsection{Fewell Case (2)}

\citet{fewell:2006} does not discuss in any detail how to compute the area of 
overlap for Fewell case (2), but this is simpler than Fewell case (1). 
In Fewell case (1) we had a circular triangle, whereas case (2) presents a 
circular quadrilateral. Before we consider the various scenarios, 
let us define under what conditions Fewell case (2) is satisfied.

For Fewell case (2), the constraint is that the two intersection points of 
circles 1 and 2 both lie outside of circle 3 (this narrows it down to Fewell 
case (2) or (7)). Therefore we must satisfy conditions $\bar{A}$ and $\bar{B}$.

Next, the intersection points of circles 1 and 3 must lie inside circle 2 (this 
distinguishes it from Fewell case (7)). Assuming case 14, condition $\bar{A}$ 
and condition $\bar{B}$, it is not physically possible to have a solution where 
one of these intersection points lies inside and the other outside of circle 2. 
It is, however, possible to have both outside, which gives rise to Fewell case 
(7). Because of this fact, we only need to use one of the intersection points 
to define condition D:

\begin{align}
&\mathbf{Condition\,\,D} \nonumber \\
&(\mathcal{X}_{13} - S_{P*})^2 + \mathcal{Y}_{13}^2 < p^2
\label{eqn:conditionD}
\end{align}

But, even this is not enough for a unique solution. In Fewell case (2) of 
Figure~5 of \citet{fewell:2006}, one could imagine mirroring the moon along the 
line connecting the two moon-star intersection points. This would still satisfy 
all of the above conditions but now no part of the moon is actively transiting. 
The nominal case, as shown in Figure~5 of \citet{fewell:2006}, is labelled as 
14.2a (shown in Figure~\ref{fig:subs}), whereas the alternative case is 14.2b. 
Therefore, we define condition E, which if satisfied gives case 14.2a and if 
anti-satisfied gives case 14.2b:

\begin{align}
&\mathbf{Condition\,\,E} \nonumber \\
&(S_{S*}-s)<(S_{P*}-p)
\label{eqn:conditionE}
\end{align}

Case 14.2a has 
$A_{S,\mathrm{transit}}^{\mathcal{E}14.2a} = \pi s^2 - \alpha_{SP}$, otherwise 
for 14.2b we have $A_{S,\mathrm{transit}}^{\mathcal{E}14.2b} = 0$.

\subsubsection{Fewell Case (3)}

To meet Fewell case (3), both intersection points of 1 and 2 must lie inside 
circle 3 i.e. condition A and B. Sub-case 14.3 contains two possible 
sub-sub-cases; 14.3a and 14.3b. The second sub-sub-case was identified quite 
late in our analysis and noticed during testing and debugging.

The two cases are differentiated by whether the planet's centre is inside or 
outside the stellar disc. If outside, then we have case 14.3a, which satisifes 
condition F.

\begin{align}
&\mathbf{Condition\,\,F} \nonumber \\
&S_{P*}>1
\label{eqn:conditionF}
\end{align}

In this case, 14.3a, 
$A_{\mathrm{S,transit}}^{\mathcal{E}14.3a} = \alpha_{S*} - \alpha_{P*}$. For 
case 14.3b, the planet is inside and we have 
$A_{\mathrm{S,transit}}^{\mathcal{E}14.3b} = \pi p^2 - \alpha_{SP} - \alpha_{P*} 
+ \alpha_{S*}$.

\subsubsection{Fewell Case (7)}

For Fewell case (7), we must satisfy conditions $\bar{A}$ and $\bar{B}$, as did 
Fewell case (2). The final condition for Fewell case (7) is simply the opposite 
of condition D, i.e. condition $\bar{D}$.

However, this is insufficient to give a unique solution. The standard Fewell 
case (7) has no interaction between the planet and moon within the stellar disc 
but this is not a condition, merely an artifact of the figure's construction. 
It is possible that the intersection points of 1 and 2 lies outside circle 3, 
\textbf{and} the intersection points of 1 and 3 lies outside 2, but still two 
distinct cases exist. These two sub-sub-cases are that if the intersection 
of 2 and 3 lies within 1, then areal interactions must be occurring. We label 
this case 14.7b. Case 14.7a is that the circles do not interact.

This first case occurs when both intersection points of 2 and 3 lie outside of 
circle 1. It is not possible to have one inside and one outside under the 
previous conditions so far met. Thus we have condition G, which if satisfied 
yields case 14.7b and if anti-satisfied gives 14.7a.

\begin{align}
&\mathbf{Condition\,\,G} \nonumber \\
&\mathcal{X}_{23}^2 + \mathcal{Y}_{23}^2 < 1
\label{eqn:conditionG}
\end{align}

Accordingly, $A_{\mathrm{S,transit}}^{\mathcal{E}14.7a} = \alpha_{S*}$. 
Otherwise the planet-moon intersections occur within the star and thereby 
necessitating a degree of interaction (case 14.7b, satisfying condition G), for 
which $A_{\mathrm{S,transit}}^{\mathcal{E}14.7b} = \alpha_{S*} - \alpha_{SP}$.

\subsubsection{Summary of Sub-Cases}

\begin{figure*}							%
\begin{center}							%
\includegraphics[width=16.8 cm]{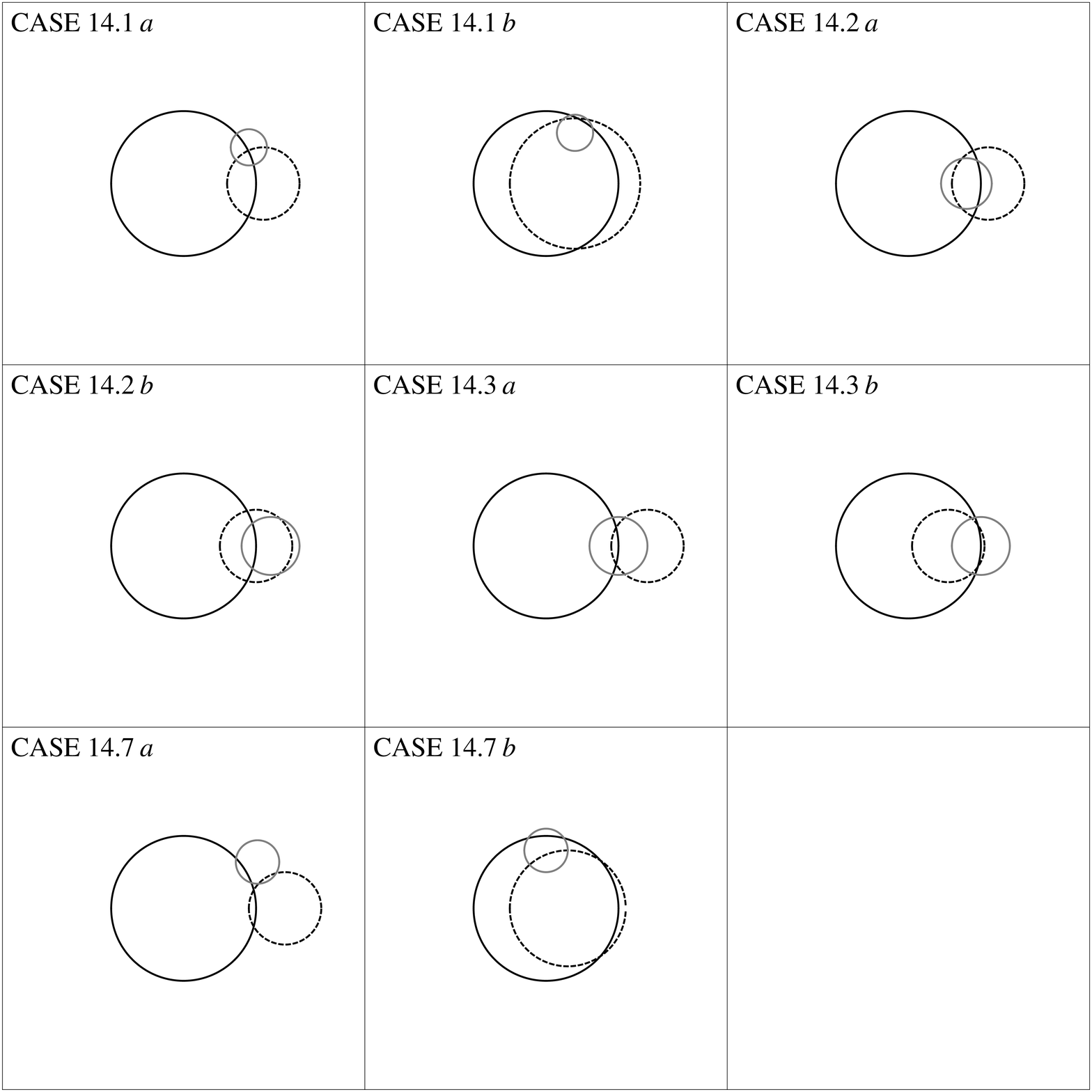}			%
\caption{\emph{Sub-cases of case 14. The subscript numbers 	%
(1,2,3,7) originate from the system employed by 		%
\citet{fewell:2006}. The extra case conditions (a,b) come from 	%
additional complexity within each sub-case yet consistent 	%
within the original case definitions of \citet{fewell:2006}.	%
The star is black solid, the planet is black dashed and the 	%
moon is gray solid. Relative sizes selected purely for 		%
depiction purposes.}} 						%
\label{fig:subs}						%
\end{center}							%
\end{figure*}

This completes every possible scenario for three circles interacting. The 
sub-case conditions are summarized in Table~\ref{tab:subs} and illustrated
example configurations are shown in Figure~\ref{fig:subs}. We 
have calculated the area of overlap in every case in an analytic manner. This 
will allow for extremely expedient computation of the light curve from a planet 
plus moon.

Due to the plethora of cases, sub-cases and even sub-sub-cases, a large amount 
of testing of the \luna\ algorithm has been executed. This was done by trying
various system configurations and ensuring no gaps existed in the case 
conditions or unusual light curve features were produced by the algorithm. It 
was during this stage which we identified that sub-case 14.3 actually had two
sub-sub-cases in the form of 14.3a and 14.3b.

\begin{table}									%
\caption{List of sub-cases and conditions. A bar indicates the condition is 	%
anti-satisfied.}								%
\centering 
\begin{tabular}{c c c} 
\hline\hline 
Sub-Case & Conditions Satisfied & $A_{\mathrm{S,transit}}$ \\ [0.5ex] 		%
\hline 
14.1a & A; $\bar{B}$; C & $\alpha_{S*} - 	
A_{\mathrm{overlap}}^{\mathcal{E}14.1a}$ \\					%
14.1b & A; $\bar{B}$; $\bar{C}$ & $\alpha_{S*} - 
A_{\mathrm{overlap}}^{\mathcal{E}14.1b}$ \\					%
14.2a & $\bar{A}$; $\bar{B}$; D; E & $\pi s^2 - \alpha_{SP}$ \\			%
14.2b & $\bar{A}$; $\bar{B}$; D; $\bar{E}$ & 0 \\				%
14.3a & A; B; F & $\alpha_{S*} - \alpha_{P*}$ \\				%
14.3b & A; B; $\bar{F}$ & $\pi p^2-\alpha_{P*}-\alpha_{SP}+\alpha_{S*}$ \\	%
14.7a & $\bar{A}$; $\bar{B}$; $\bar{D}$; $\bar{G}$ & $\alpha_{S*}$ \\		%
14.7b & $\bar{A}$; $\bar{B}$; $\bar{D}$; G & $\alpha_{S*} - \alpha_{SP}$ \\[1ex]%
\hline\hline 
\end{tabular}									%
\label{tab:subs} 
\end{table}									%

\subsection{Flow Decision Diagram}

With all the conditions stated, it is useful to construct a flow diagram of the 
decision route which \luna\ should take. This flow diagram should be designed to 
minimize the number of calculations \luna\ has to perform. We have conditions A 
to G for the sub-cases and three additional conditions from the principal cases, 
so the maximum number of decisions in any chain should be 10. Ideally, most 
decision routes will involve fewer than 10 decisions.

\subsubsection{Decision 1}

The best-place to start is with the principal cases. The computation of the 
transit light curve will always require a computation of $S_{P*}$, as it is 
assumed that the planet always transits.

Since we always have to compute this term anyway, the economic way to proceed is 
to base a decision tree upon its behaviour. This essentially gives us a three 
way-split into three decision trees: cases 1 through 9 (called ``p-out''), 10 
through 19 (``p-part'') and 20 through 27 (``p-in'').  

\subsubsection{Decision 2}

Having computed $S_{P*}$, should we compute $S_{S*}$ and $S_{PS}$? In several 
cases, only one of these is required to calculate the light curve. Consider, the 
p-out tree first (note that the second decision does not have to be the same in 
each tree). Since the planet is out of the stellar disc, areal interaction is 
impossible and thus $S_{PS}$ need never be computed. Therefore, the logical 
second decision for p-out is simply to evaluate $S_{S*}$ and see which of 
the three possible circumstances it is in.

For the p-in tree, there are two possible questions one could pose. Remember, we 
wish to pose the question which leads to the most efficient algorithm. First 
possible question: evaluate $S_{S*}$ and see which state it is in. Second 
possible question: evaluate $S_{SP}$ and see which state it is in.

If we ask the second question, the in-case requires no further computations. 
However, the part and out cases both require that we evaluate $S_{S*}$ to 
proceed.

If we ask the first question, the out case requires no further computation. 
However, the in and part cases require we calculate $S_{SP}$. Therefore, the two 
approaches seem to yield equivalent levels of complexity. In general, if the 
moon is transiting, $S_{S*}$ will always be needed whereas $S_{SP}$ is only 
needed if areal interaction occurs. Therefore, we prefer to ask about $S_{S*}$ 
first.

Finally, for the p-part tree, case 14 is the mix which is by the far the most 
complicated case computationally. We wish to choose a tree based upon minimizing 
our exposure to case 14. $S_{S*}$ again makes a natural choice by the groupings 
of the actively transiting area formulas.

\subsubsection{Decision 3}

Since decision 2 is always the same, decision 3 is simply to evaluate $S_{SP}$ 
and check its condition. There is no choice in this.

\subsubsection{Decision 14.1}

If case 14 occurs, we have a new decision tree opening up due to presence of 
sub-cases. The first decision is decision 14.1. Unlike the principal tree, each 
sub-case does not necessarily involve a calculation of all the possible 
conditions. Therefore, some conditions crop up more frequently than others. If 
all the conditions took the same time to compute, then we would prefer to 
compute those conditions which occur most frequently early on in our decision 
tree.

Conditions A and B clearly provide the backbone to the decision tree and thus 
must be evaluated early on. Their evaluation is ultimately unavoidable so should 
be done at the start. Case 14.3 is unusual in that it can be identified as 
satisfying condition B only since all sub-cases anti-satisfy B. Condition B is 
therefore the obvious starting point in our decision tree. If verified, we have 
case 14.3 and if not we move into decision 14.2.

\subsubsection{Decision 14.2}

Since conditions A and B are the backbone and B has been evaluated, the next 
step is to evaluate A. If satisfied, we have case 14.1. Anti-satisfaction means 
we have either 14.2 or 14.7.

\subsubsection{Decision 14.3}

If we have 14.1, then decision 14.3 must be to evaluate condition C. If we have 
14.2 or 14.7, then the discriminator is condition D.

\subsubsection{Decision 14.4}

Not needed for 14.1 cases, but for Fewell cases (2) and (7) we must compute 
conditions E and G respectively.

\subsubsection{Summary of decision tree}

\begin{figure*}							%
\begin{center}							%
\includegraphics[width=16.8 cm]{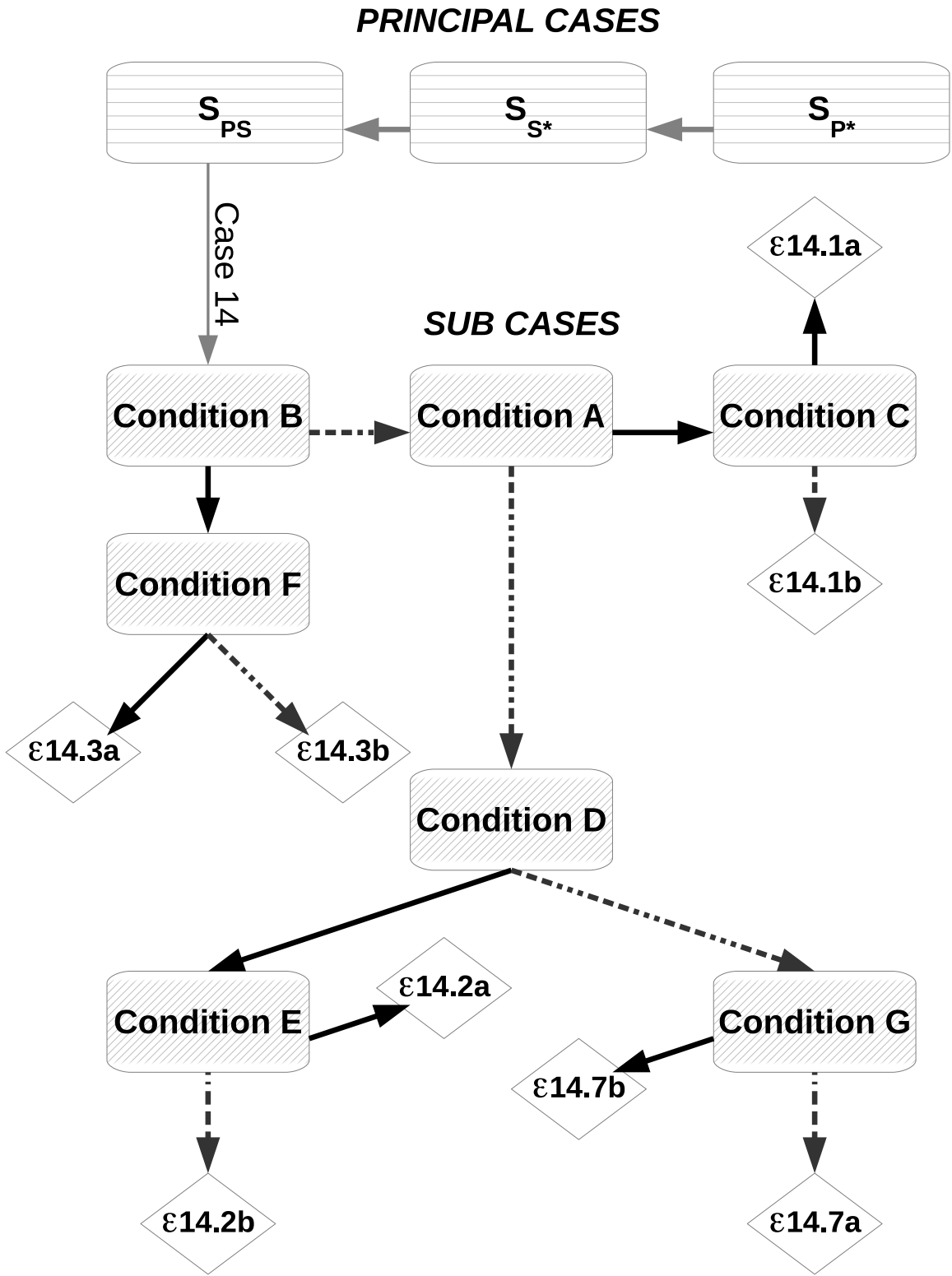}		%
\caption{\emph{Flow chart showing the decision process of	%
choosing which case we are in. The decision tree has been	%
optimized to marginalize the more CPU intensive computations	%
until absolutely required.					%
The principal cases are not					%
shown but are easily seen in Table~\ref{tab:principals}.	%
Square boxes indicate an evaluation of whatever is inside	%
the box.							%
Black arrows indicate a true statement and the dot-dashed	%
gray arrows indicate a false.}}					%
\label{fig:decisions}						%
\end{center}							%
\end{figure*}

Figure~\ref{fig:decisions} shows the decision process in a flow chart to 
summarize the approach of \luna.

\section{Example Transit Light Curves}
\label{sec:examples}

\subsection{Dealing with Inclination}

Transit modellers will be familiar with the pot-holes in the road presented when 
trying to fit transit light curves with the physical parameters. To start with, 
inclination is almost always in the range $\sim$85 to 90 degrees and fitting for 
$i_{B*}$ yields large correlations and lethargic routines. In practice, it is 
better to define the impact parameter of the transit chord across the star, 
$b_{B*}$.

\begin{equation}
b_{B*} = (r_{B*}/R_*) \cos i_{B*}
\end{equation}

A similar strategy would seem advisable for moons and we can define an analogous
quantity:

\begin{equation}
b_{SB} = (r_{SB}/R_P) \cos i_{SB}
\end{equation}

It should be noted that $b_{SB}$ is only the impact parameter in the reference 
frame of the moving barycentre and will not be the observed impact parameter in 
the sky frame. Therefore, $b_{SB}$ can be greater than unity and yet the moon
still transits the star. 

Another subtlety is that $i_{SB} = \pi/2 + \delta$ would give
a different light curve than $i_{SB} = \pi/2 - \delta$, but both would yield the
same value for $b_{SB}$ (in contrast to the situation for a planet by itself). 
To account for this asymmetry, negative impact parameters are perceived as
yielding $i_{SB} = \pi - \cos^{-1}[R_P b_{SB}/r_{SB}]$. Retrograde satellites
with inclinations in the range $\pi<i_{SB}<2\pi$ are adjusted in a second step
where if a logical flag for retrograde orbits is switched on, $\pi$ is added
onto the inclination.

\subsection{Fitting Parameter Set}

A typical planetary parameter set would be $\{\tau,p^2,(a/R_*),b\}$.
Due the strong correlation between $(a/R_*)$ and $b$ \citep{carter:2008}, it
is preferable to switch one of these parameters for a less correlated term.
A typical choice is $\tilde{T}$, the duration for the planet to move from its
centre overlapping the stellar limb to exiting under the same condition.
An inverse-mapping expression to go from $\tilde{T}$ to $(a/R_*)$ is given in
\citet{investigations:2010}. Note that a $\tilde{T}$ equation for the moon is
not possible since $b_{SB}>1$ is permitted and thus $\tilde{T}$ would be 
complex.

We now have the four fitted parameters $\{\tau_{B*},p^2,\tilde{T}_{B*},b_{B*}\}$ 
for the planet-moon barycentre. Additionally, we need the orbital period, 
$P_{B*}$, which may be fitted for if multiple transits exist. Radial velocity or 
secondary eclipse information may be used to expand the parameter set to include 
$e_{B*} \cos \omega_{B*}$ and $e_{B*} \sin \omega_{B*}$. Finally, the 
out-of-transit flux, OOT, and the blending factor, $B$ (which cannot be fitted
for) should also be included. The quadratic limb darkening coefficients
$u_1$ and $u_2$ give an additional two parameters bringing the total to 11.
RV data may also require additional terms such as semi-amplitude, $K$, velocity
offset, $\gamma$, etc.

To include the exomoon, 9 new parameters are needed: 
$\{\phi_{SB}$, $(M_S/M_P)$, $s$, $(a_{SB}/R_*)$, $b_{SB}$, $P_{SB}$, 
$e_{SB}\cos\omega_{SB}$, $e_{SB}\sin\omega_{SB}$, $\Omega_{SB}\}$.
In total, this means we have 20+RV parameters. In practice, not all of these
will be fitted for e.g. the blending factor.

\subsection{Example Simulations \& Fits}
\label{sub:exampledetails}

In the next three subsections (\S\ref{sub:NepM2close}-\ref{sub:NepM2none}) we 
provide three simulations generated by \luna. The system configuration used in 
each case is discussed in the relevant subsections. In each case, we use a 
cadence of 1\,minute and add on Gaussian noise of 250\,ppm, in-line with the 
properties of \emph{Kepler}'s short-cadence photometry 
\citep{kippingbakos:2011b}. Data are produced surrounding $\pm0.5$\,days of 
$\tau_{B*}$ for $N$ consecutive transit epochs.

These noised light curves are then fitted
using a Metropolis-Hastings MCMC routine with 125,000 steps and a 20\% burn-in
to give $10^5$ final points for building the parameter posteriors.
Jump-sizes are set to be equal to the 1-$\sigma$ uncertainties (determined
through preliminary runs) and the final parameter values are given by the 
median of each parameter with uncertainties of 34.15\% quantiles either side.

For each case, we perform three fits with fixed assumptions i) no moon is
present i.e. $s=0$ and $M_S/M_P=0$ ii) the moon is prograde (i.e. $i_{SB}$ is
bounded to be in the range $0<i_{SB}\leq\pi$) iii) the moon is retrograde
($\pi<i_{SB}\leq2\pi$). In reality, only one case is genuine and the two
other act to show how distinguishable each scenario really is. We always assume
a circular orbit for the moon and planet for simplicity, which removes four
parameters.

In all cases, we utilize a powerful trick pointed out in \citet{weighing:2010}.
The detection of a planet-moon system allows one to determine the absolute
mass and radius of the host star through Kepler's Third Law alone. Armed with
these parameters, the planetary and lunar physical dimensions may also be
obtained, thus replacing the traditional need for invoking spectroscopy and 
stellar evolution models. The only additional information one requires is the
radial velocity semi-amplitude induced by the planet-moon barycentre, $K$. As 
the observational uncertainty on this parameter scales with 
$\sim1/\sqrt{N_{\mathrm{RV}}}$, where $N_{\mathrm{RV}}$ is the number of RV 
measurements, it can be determined to high precision by simply repeating the 
radial velocity observations. 

As shown in \citet{weighing:2010}, the absolute dimensions of the star have 
scalings $M_* \sim K_*^3$ and $R_* \sim K_*$, and so
$[\sigma(M_*)]^2 \simeq [3\sigma(K_*)]^2 + [\sigma(M_*')]^2$ and 
$[\sigma(R_*)]^2 \simeq [\sigma(K_*)]^2 + [\sigma(R_*')]^2$, where
the dashed terms denote the value derived assuming no error on $K_*$. In an 
example shown later (see \S\ref{sub:NepM2far}), $K_* = 6$\,m/s and 
$\sigma(M_*')/M_*'\simeq15\%$ and $\sigma(R_*')/R_*'\simeq6\%$. This example
is a singular case but nonetheless serves the purpose of illustrating a feasible 
situation. In this example, the assumption that $K_*$ contributes negligible 
error into the stellar mass determination is valid for 
$\sigma(K_*)/K_*\lesssim$2.5\% ($\sim0.15$\,m/s) and for the stellar radius 
$\sigma(K_*)/K_*\lesssim$7.0\% ($\sim0.4$\,m/s). These RV precisions are 
comparable to values being reported by those using high-precision facilities 
such as Keck and HARPS \citep{vogt:2010,lovis:2011}. More generally, once an 
exomoon system is found, the uniqueness and importance of the system would 
certainly warrant the exploitation of such spectrographic resources and so high 
levels of precision in $K_*$ can be expected.

We therefore assume that the uncertainty on $K_*$ is 
much less than the uncertainty on the photometrically determined exomoon 
parameters in what follows. The inclusion of the determination of $M_*$ and 
$R_*$ using the \citet{weighing:2010} method will demonstrate the feasibility of 
the technique.

We note that $s$ and $M_S/M_P$ are positive definite and thus always yield a 
non-zero result. To identify cases where this bias is creating a false-positive, 
we apply the \citet{lucy:1971} test. If $s$ or $M_S/M_P$ yield a 
false-alarm-probability below 5\%, then we quote the MCMC result as usual. 
Otherwise, we quote the 95\% confidence upper limit on those parameters.

\subsection{HZ-Neptune with a Close, Prograde Moon around an M2 Star}
\label{sub:NepM2close}

\subsubsection{Simulation and Fitting}

As a first example, we consider a Neptune-mass and radius planet in orbit
an M2 star in the habitable-zone ($K_*=6.0$\,m\,s$^{-1}$). Using the stellar 
parameters from \citet{cox:2000}, we assign $M_* = 0.40$\,$M_{\odot}$ and 
$R_* = 0.50$\,$R_{\odot}$. We place the planet in an orbital period of
$P_{B*}=46.0$\,days in a circular orbit with $i_{B*} = 90^{\circ}$. Quadratic 
limb darkening coefficients were generated using a \citet{kurucz:2006} style
atmosphere following the methodology of \citet{kippingbakos:2011b}, giving
$u_1 =0.3542$ and $u_2 = 0.3607$. In total, we generate $N=6$ transit epochs 
spanning 0.85\,years of continuous photometry.

For the exomoon, we consider an Earth-mass and radius moon on a close, prograde
orbit around the host planet. We therefore select $a_{SB}$ to be 5\% of the Hill
radius, corresponding to $\simeq2$\,$R_P$. Through Kepler's Third Law, the
orbital period is $P_{SB} = 0.3142$\,days. Such a moon would be experiencing
large tidal dissipation and may not necessarily persist for Gyr, however, this
is irrelevant for the purpose of generating some example transit light curves. 
We choose to place the moon in a slightly inclined orbit of 
$i_{SB} = 92.93^{\circ}$, corresponding to $b_{SB} = -0.1$. We also use 
$\Omega_{SB} = 5^{\circ}$ and assume a circular orbit. In the left panel of 
Figure~\ref{fig:NepM2close_big}, the simulated light curve is shown, before any 
noise is added.

Due to the low impact parameter of the barycentre and the high coplanarity,
TDV-TIP effects will be small and thus a determination of the sense of
orbital motion will be highly challenging (see \S\ref{sub:retrograde} for
explanation). However, coplanarity is a reasonable choice as highly inclined
moons have reduced Hill stability \citep{donnison:2010} and thus are less
likely a-priori. The same argument is true for eccentric moons 
\citep{domingos:2006}.

\subsubsection{Results}

As discussed in \S\ref{sub:exampledetails}, we performed three fits for
prograde, retrograde and no moon assumptions. A comparison of the fitted
parameters for each assumption is presented in Table~\ref{tab:NepM2close},
including the $\chi^2$ and Bayesian Information Criterion (BIC) of the best-fits 
for each model. BIC \citep{schwarz:1978,liddle:2007} is a tool for model
selection which severely penalizes models for including more degrees of freedom
(i.e. Occam's razor) and is defined by:

\begin{equation}
\mathrm{BIC} = \sum_{i=1}^N \Bigg( \frac{ f_{i}^{\mathrm{obs}} - 
f_{i}^{\mathrm{model}} }{ \sigma_i } \Bigg)^2 + k \ln N
\label{eqn:BICeqn}
\end{equation}

where $f_i$ denotes flux, $\sigma_i$ is the associated uncertainty, $N$ is the
number of observations and $k$ is the number of free parameter in the fit.
We find that the prograde moon is only marginally preferable to 
the retrograde model, as expected. However, the no-moon model is clearly
a poor fit and would be rejected. An F-test finds the prograde moon model
accepted over no-moon model with a confidence of 24.0-$\sigma$, representing
a very secure detection.

\begin{figure*}							
\begin{center}							
\includegraphics[width=16.8 cm]{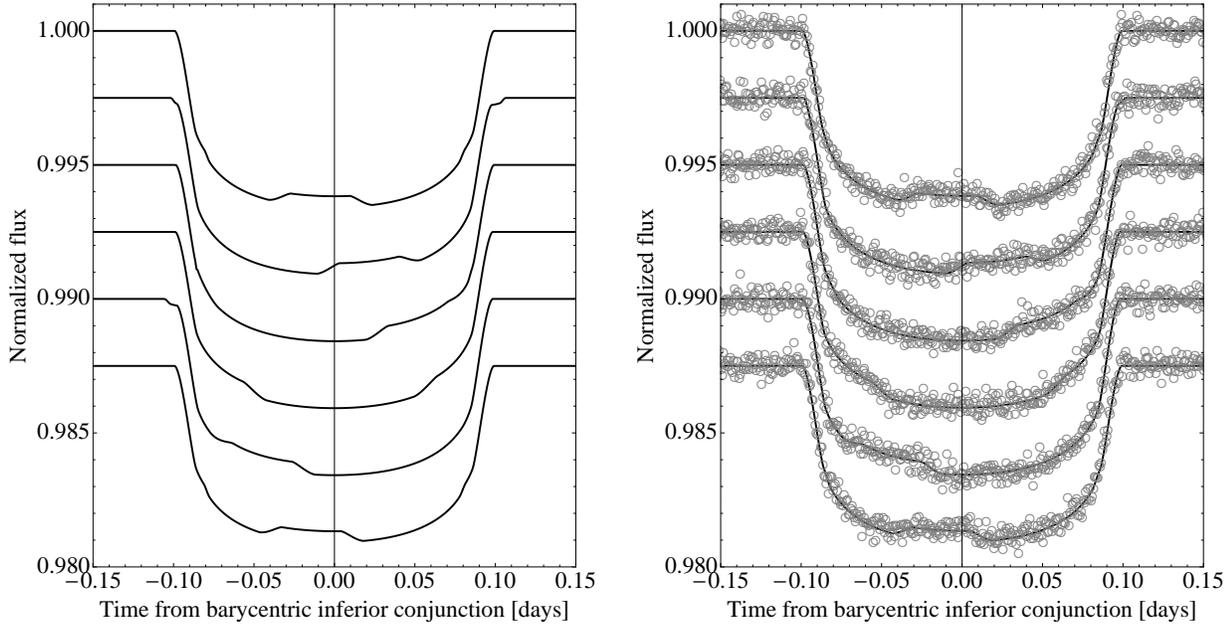}			
\caption{\emph{Left panel: Simulation from \luna\ of a habitable-zone Neptune
with a close, prograde, Earth-like moon for an M2 host star. Right panel:
Noised data (circles) of 250\,ppm per minute overlaid with best fit from an
MCMC routine (solid).}}		
\label{fig:NepM2close_big}						
\end{center}							
\end{figure*}

The fit assuming a prograde orbit is shown in the right panel of 
Figure~\ref{fig:NepM2close_big}, where the thick black line represents the 
actual best-fit (i.e. \emph{not} the original simulation light curve).

\begin{table*}
\caption{\emph{Comparison of parameter estimates from various model assumptions
used in the fits. Data generated for a Neptune with a close moon around an M2 
star.}} 
\centering 
\begin{tabular}{c c c c c} 
\hline\hline 
\textbf{Parameter} & \textbf{Truth} & \textbf{Prograde} & \textbf{Retrograde} 
& \textbf{No Moon} \\ [0.5ex] 
\hline
$\chi^2$ & 8786.35 & 8773.48 & 8775.23 & 9414.23 \\
BIC & - & 8891.31 & 8893.07 & 9468.62 \\
\hline
\emph{Fitted params.} & & & & \\
\hline 
$P_{B*}$ [days] & $46.000000$ & $45.999997_{-0.000075}^{+0.000077}$ & 
$46.000000_{-0.000078}^{+0.000080}$ & $45.999999_{-0.000085}^{+0.000085}$\\
$\tau_{B*}$ [BJD - 2454000] & $956.00000$ & $956.00001_{-0.00024}^{+0.00023}$ & 
$956.00000_{-0.00024}^{+0.00024}$ & $955.99996_{-0.00026}^{+ 0.00026}$ \\
$p^2$ [\%] & $0.5071$ & $0.5082_{-0.0041}^{+0.0042}$ & 
$0.5082_{-0.0042}^{+0.0042}$ & $0.5306_{-0.0027}^{+0.0038}$ \\
$b_{B*}$ & $0.00$ & $-0.02_{-0.14}^{+0.15}$ & $0.02_{-0.14}^{+0.14}$ & 
$0.01_{-0.17}^{+0.16}$ \\
$\tilde{T}_{B*}$ [s] & $15882$ & $15877_{-26}^{+26}$ & $15877_{-26}^{+26}$ & 
$15895_{-26}^{+26}$ \\
$P_{SB}$ [days] & $0.3142001$ & $0.3142029_{-0.0000066}^{+0.0000067}$ & 
$0.3142026_{-0.0000068}^{+0.0000069}$ & - \\
$\phi_{SB}$ [$^{\circ}$] & $40$ & $71_{-40}^{+24}$ & $21_{-14}^{+12}$ & - \\
$s$ & $0.0183$ & $0.0181_{-0.0011}^{+0.0011}$ & $0.0182_{-0.0011}^{+0.0012}$ & 
$0.0000$ \\
$a_{SB}/R_*$ & $0.139$ & $0.134_{-0.014}^{+0.012}$ & $ 0.129_{-0.015}^{+0.013}$ 
& - \\
$b_{SB}$ & $-0.1$ & $-0.21_{-0.95}^{+1.25}$ & $0.04_{-0.64}^{+0.58}$ & - \\
$\Omega_{SB}$ [$^{\circ}$] & $5$ & $34_{-39}^{+30}$ & $-76_{-12}^{+14}$ & - \\
$M_{S}/M_{P}$ & $0.058$ & $0.073_{-0.020}^{+0.027}$ & 
$0.0694_{-0.018}^{+ 0.019}$ & $0.000$ \\ 
\hline
\emph{Physical params.} \\
\hline
$M_*$ [$M_{\odot}$] & $0.40$ & $0.45_{-0.23}^{+0.66}$ & $0.69_{-0.39}^{+1.27}$ 
& - \\
$R_*$ [$R_{\odot}$] & $0.50$ & $0.52_{-0.11}^{+0.19}$ & $0.61_{-0.15}^{+0.25}$ 
& - \\
$M_P$ [$M_J$] & $0.054$ & $0.057_{-0.022}^{+0.047}$ & $0.077_{-0.032}^{+0.076}$ 
& - \\ 
$R_P$ [$M_J$] & $0.346$ & $0.363_{-0.078}^{+0.128}$ & $0.42_{-0.10}^{+0.17}$ 
& - \\
$M_S$ [$M_{\oplus}$] & $1.00$ & $1.31_{-0.58}^{+1.34}$ & $1.64_{-0.79}^{+2.08}$ 
& - \\
$R_S$ [$R_{\oplus}$] & $1.00$ & $1.03_{-0.22}^{+0.38}$ & $1.19_{-0.29}^{+0.54}$ 
& - \\
$\rho_S$ [g\,cm$^{-3}$] & $5.5$ & $6.5_{-2.2}^{+3.0}$ & $5.2_{-1.8}^{+2.2}$ 
& - \\ [1ex]
\hline\hline 
\end{tabular}
\label{tab:NepM2close} 
\end{table*}

We find that all of the parameters from the prograde fit are consistent with the
true model. The physical parameters are generally poorly constrained, but
apparently sufficient to identify the moon as rocky rather than icy.

\subsection{HZ-Neptune with a Far, Retrograde Moon around an M2 Star}
\label{sub:NepM2far}

\subsubsection{Simulation and Fitting}

Our second example is identical to the previous one in \S\ref{sub:NepM2close},
except now we push the orbiting moon into a more distant, retrograde orbit.
We place the moon at 90\% of the Hill radius (and thus inside the 93.09\% limit
determined by \citet{domingos:2006}).

\subsubsection{Results}

A comparison of the fitted parameters for each of our three model assumptions is 
presented in Table~\ref{tab:NepM2far}, including the $\chi^2$ and BIC values of 
the best-fits for each model. We find that the retrograde moon is only 
marginally preferable to the prograde model, as expected. However, the no-moon 
model is clearly a poor fit and would be rejected. An F-test finds the 
retrograde moon model accepted over no-moon model with a confidence of 
50.3-$\sigma$, representing a very secure detection.

\begin{figure*}							
\begin{center}							
\includegraphics[width=16.8 cm]{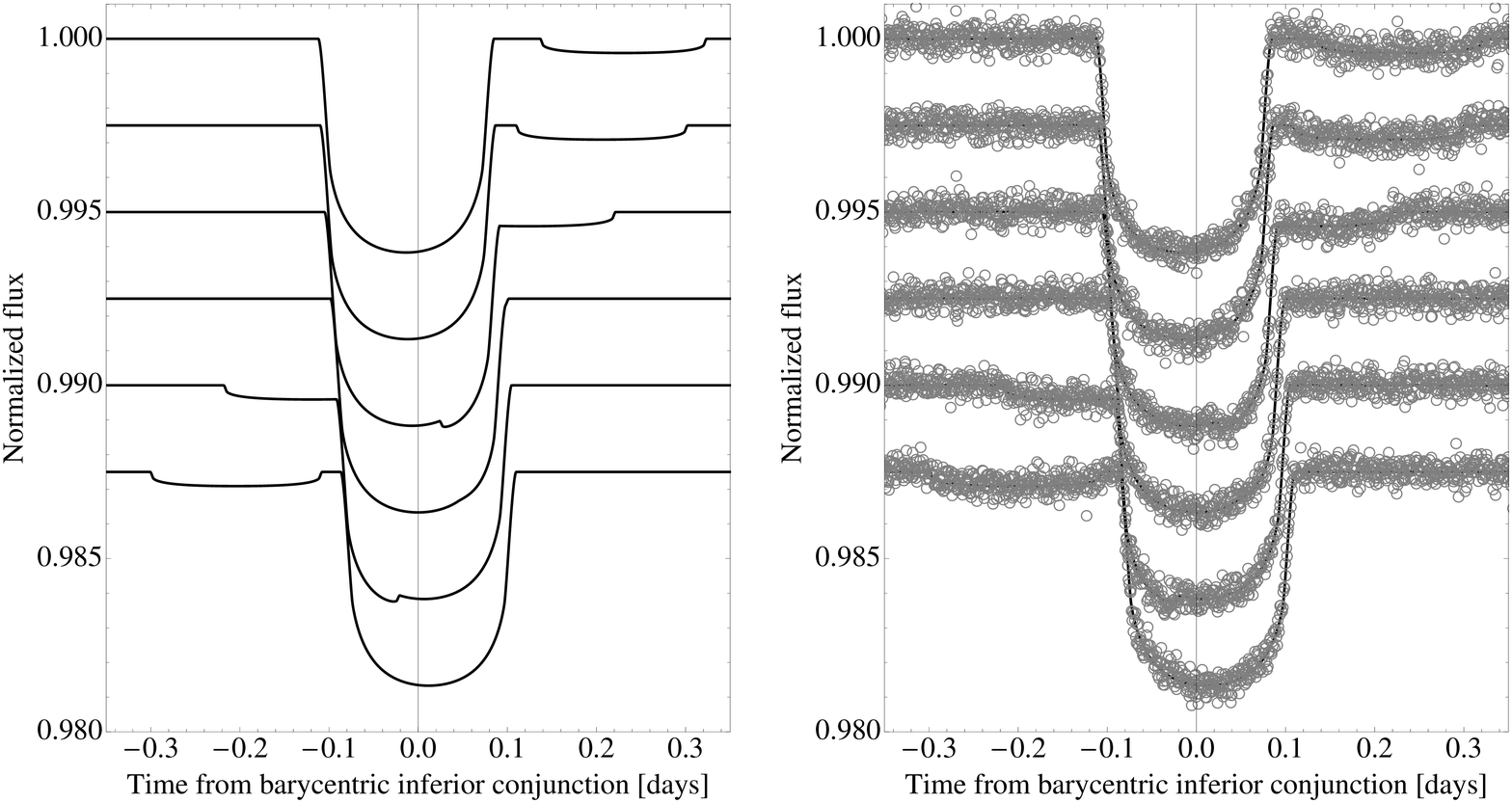}			
\caption{\emph{Left panel: Simulation from \luna\ of a habitable-zone Neptune
with a distant, retrograde, Earth-like moon for an M2 host star. Right panel:
Noised data (circles) of 250\,ppm per minute overlaid with best fit from an
MCMC routine (solid).}}		
\label{fig:NepM2far_big}						
\end{center}							
\end{figure*}

The fit assuming a retrograde orbit is shown in the right panel of 
Figure~\ref{fig:NepM2far_big}, where the thick black line represents the 
actual best-fit (i.e. \emph{not} the original simulation light curve).

\begin{table*}
\caption{\emph{Comparison of parameter estimates from various model assumptions
used in the fits. Data generated for a Neptune with a distant moon around an M2 
star.}} 
\centering 
\begin{tabular}{c c c c c} 
\hline\hline 
\textbf{Parameter} & \textbf{Truth} & \textbf{Retrograde} & \textbf{Prograde} & 
\textbf{No Moon} \\ [0.5ex] 
\hline
$\chi^2$ & 8529.95 & 8518.65 & 8518.68 & 11477.45 \\
BIC & - & 8636.48 & 8636.51 & 11531.83 \\
\hline
\emph{Fitted params.} & & & & \\
\hline 
$P_{B*}$ [days] & $46.000000$ & $45.99964_{-0.00064}^{+0.00061}$ & 
$45.99967_{-0.00072}^{+0.00058}$ & $46.005178_{-0.000085}^{+0.000084}$ \\
$\tau_{B*}$ [BJD - 2454000] & $956.00000$ & $956.0010_{-0.0018}^{+0.0019}$ & 
$956.0010_{-0.0017}^{+0.0021}$ & $955.98483_{-0.00026}^{+0.00026}$ \\
$p^2$ [\%] & $0.5071$ & $0.5088_{-0.0023}^{+0.0029}$ & 
$0.5088_{-0.0023}^{+0.0029}$ & $0.5075_{-0.0028}^{+0.0044}$ \\
$b_{B*}$ & $0.00$ & $0.02_{-0.14}^{+0.13}$ & $0.01_{-0.14}^{+0.14}$ & 
$0.00_{-0.19}^{+0.19}$ \\
$\tilde{T}_{B*}$ [s] & $15882$ & $15896_{-26}^{+26}$ & $15896_{-26}^{+26}$ & 
$15883_{-27}^{+27}$ \\
$P_{SB}$ [days] & $23.995$ & $23.984_{-0.031}^{+0.031}$ & 
$23.987_{-0.034}^{+0.029}$ & - \\
$\phi_{SB}$ [$^{\circ}$] & $40$ & $34_{-29}^{+70}$ & $143_{-77}^{+46}$ & - \\
$s$ & $0.01834$ & $0.01842_{-0.00049}^{+ 0.00048}$ & 
$0.01843_{-0.00049}^{+0.00048}$ & $0.00000$ \\
$a_{SB}/R_*$ & $2.509$ & $2.481_{-0.044}^{+0.045}$ & $2.478_{-0.044}^{+0.044}$ 
& - \\
$b_{SB}$ & $-0.1$ & $-0.02_{-0.83}^{+0.61}$ & $-0.09_{-0.59}^{+0.57}$ & - \\
$\Omega_{SB}$ [$^{\circ}$] & $5$ & $11_{-70}^{+29}$ & $7_{-77}^{+48}$ & - \\
$M_{S}/M_{P}$ & $0.0583$ & $0.0622_{-0.0070}^{+0.0076}$ & 
$0.0619_{-0.0067}^{+0.0084}$ & $0.0000$ \\ 
\hline
\emph{Physical params.} \\
\hline
$M_*$ [$M_{\odot}$] & $0.400$ & $0.399_{-0.064}^{+0.061}$ & 
$0.403_{-0.066}^{+0.061}$ & - \\
$R_*$ [$R_{\odot}$] & $0.500$ & $0.504_{-0.029}^{+0.025}$ & 
$0.506_{-0.029}^{+0.025}$ & - \\
$M_P$ [$M_J$] & $0.0540$ & $0.0537_{-0.0061}^{+0.0055}$ & 
$0.0541_{-0.0063}^{+0.0055}$ & - \\ 
$R_P$ [$M_J$] & $0.346$ & $0.350_{-0.020}^{+0.018}$ & $0.351_{-0.020}^{+0.018}$ 
& - \\
$M_S$ [$M_{\oplus}$] & $1.00$ & $1.05_{-0.12}^{+0.13}$ & $1.06_{-0.12}^{+0.13}$ 
& - \\
$R_S$ [$R_{\oplus}$] & $1.000$ & $1.011_{-0.064}^{+0.059}$ & 
$1.015_{-0.065}^{+0.059}$ & - \\
$\rho_S$ [g\,cm$^{-3}$] & $5.50$ & $5.62_{-0.85}^{+1.03}$ & 
$5.58_{-0.82}^{+1.07}$ & - \\ [1ex]
\hline\hline 
\end{tabular}
\label{tab:NepM2far} 
\end{table*}

We find that all of the parameters from the retrograde fit are consistent with 
the true model. The physical parameters are generally well constrained, in
contrast to the close, prograde moon considered in \S\ref{sub:NepM2close}. This
is due to $a_{SB}/R_*$ being significantly larger and thus determined to a
higher precision which feeds into the other parameters.

\subsection{HZ-Neptune without a Moon around an M2 Star}
\label{sub:NepM2none}

\subsubsection{Simulation and Fitting}

To complete the picture, we simulate the same case as the previous two
subsections but removing the moon altogether. The data are then fitted
assuming the three models as before. The purpose of this data set is to
show what happens when \luna\ is implemented on control data and to ensure
such data does not produce false positive moon detections.

\subsubsection{Results}

A comparison of the fitted parameters for each of our three model assumptions is 
presented in Table~\ref{tab:NepM2none}, including the $\chi^2$ and BIC values of 
the best-fits for each model. We find that the no-moon model gives the lowest 
BIC by a considerable margin, as expected\footnote{Note that the $\chi^2$ is
actually worst for the no-moon model, but the BIC is lowest. This is because
BIC heavily penalizes models for including more degrees of freedom, see
Equation~\ref{eqn:BICeqn}}. The retrograde and prograde models produce 
non-sensical values for most parameters, especially the physical parameters, 
exploring various scenarios with no clear minimum in $\chi^2$ space. Note that 
$s$ and $M_S/M_P$ are positive definite and therefore are slightly skewed from a 
zero value, but not significantly so. An F-test finds the retrograde moon model 
(the best moon fit) accepted over no-moon model with a confidence of 
2.0-$\sigma$, which indicates that the detection criterion is clearly in excess 
of 2-$\sigma$. The use of BIC is therefore more accurate as a model selection
tool for exomoon detection.

\begin{figure*}							
\begin{center}							
\includegraphics[width=16.8 cm]{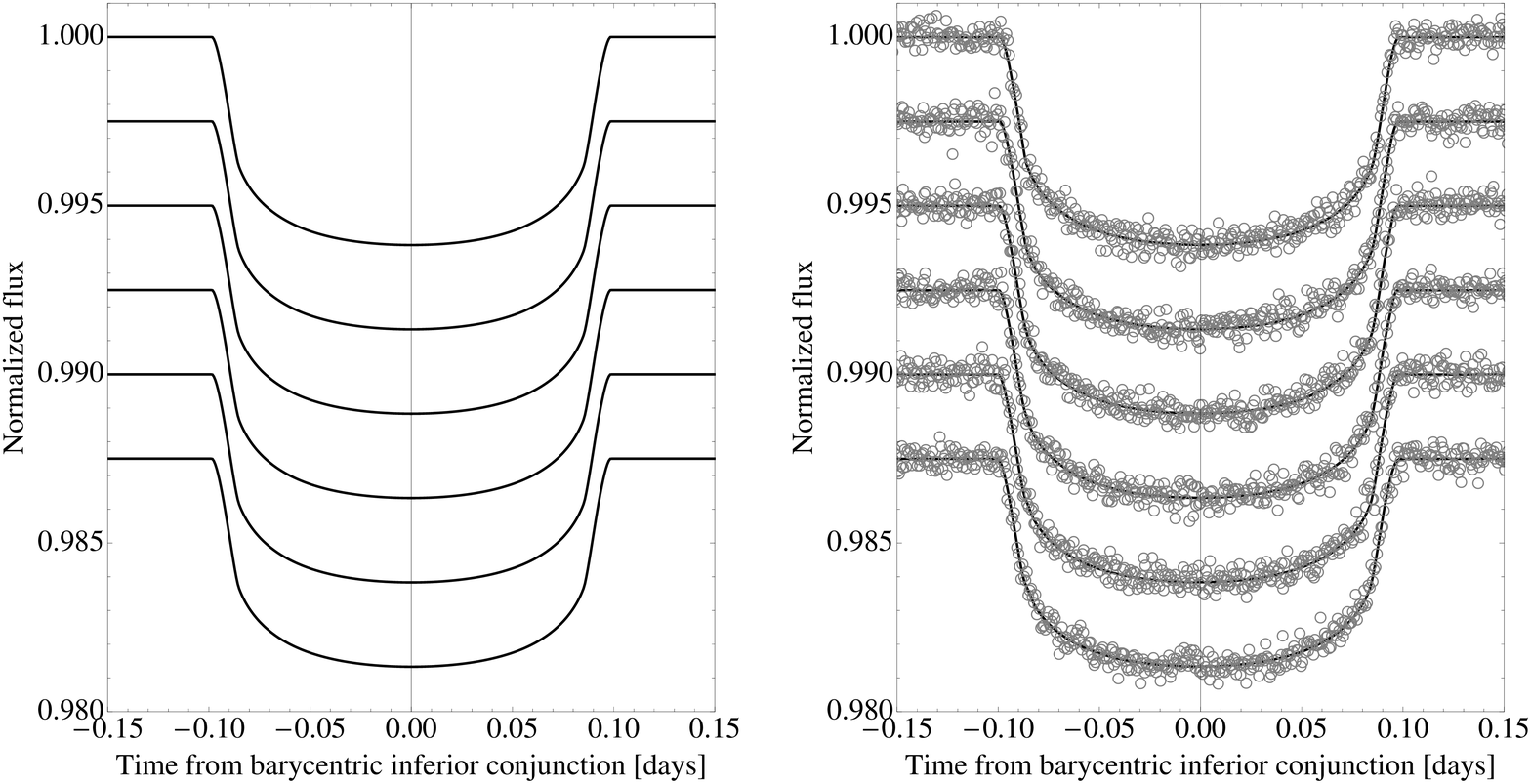}			
\caption{\emph{Left panel: Simulation from \luna\ of a habitable-zone Neptune
with no moon. Right panel: Noised data (circles) of 250\,ppm per minute overlaid 
with best fit from an MCMC routine (solid), from the starting point of assuming
a retrograde moon is in orbit.}}		
\label{fig:NepM2none_big}						
\end{center}							
\end{figure*}

\begin{table*}
\caption{\emph{Comparison of parameter estimates from various model assumptions
used in the fits. Data generated for a Neptune with no moon around an M2 star.}}
\centering 
\begin{tabular}{c c c c c} 
\hline\hline 
\textbf{Parameter} & \textbf{Truth} & \textbf{No Moon} & \textbf{Prograde} & 
\textbf{Retrograde} \\ [0.5ex] 
\hline
$\chi^2$ & 8936.23 & 8930.96 & 8928.42 & 8921.57 \\
BIC & - & 8985.34 & 9046.26 & 9039.41 \\
\hline
\emph{Fitted params.} & & & & \\
\hline 
$P_{B*}$ [days] & $46.000000$ & $46.000027_{-0.000085}^{+0.000085}$ & 
$46.00001_{-0.00048}^{+0.00057}$ & \\
$\tau_{B*}$ [BJD - 2454000] & $956.00000$ & $955.99998_{-0.00026}^{+0.00026}$ & 
$956.0000_{-0.0018}^{+0.0015}$ & $955.9992_{-0.0076}^{+0.0019}$ \\
$p^2$ [\%] & $0.5071$ & $0.5080_{-0.0026}^{+0.0035}$ & 
$0.005070_{-0.000030}^{+0.000035}$ & $0.5065_{-0.0041}^{+0.0037}$ \\
$\tilde{T}_{B*}$ [s] & $15882$ & $15883_{-27}^{+27}$ & $15881_{-33}^{+32}$ & 
$15883_{-55}^{+66}$ \\
$\phi_{SB}$ [$^{\circ}$] & - & - & $177_{-113}^{+133}$ & $223_{-94}^{+84}$ \\
$s$ & $0.0000$ & $0.0000$ & $\leq0.0094$ & $\leq0.0178$ \\
$a_{SB}/R_*$ & - & - & $1.15_{-0.57}^{+1.55}$ & $1.10_{-0.74}^{+1.86}$ \\
$b_{SB}$ & - & - & $-6.0_{-3.4}^{+4.1}$ & $-1.6_{-1.9}^{+3.4}$ \\
$\Omega_{SB}$ [$^{\circ}$] & - & - & $4_{-66}^{+59}$ & $-16_{-52}^{+79}$ \\
$M_{S}/M_{P}$ & $0.00$ & $0.00$ & $\leq0.13$ & $\leq0.37$ \\ 
\hline
\emph{Physical params.} \\
\hline
$M_*$ [$M_{\odot}$] & $0.40$ & - & $490_{-490}^{+162940}$ & 
$321_{-321}^{+5876270}$ \\
$R_*$ [$R_{\odot}$] & $0.50$ & - & $5.4_{-5.0}^{+32.0}$ & 
$4.7_{-4.4}^{+119.1}$ \\
$M_P$ [$M_J$] & $0.054$ & - & $6.3_{-6.3}^{+293.1}$ & 
$4.5_{-4.5}^{+3043.3}$ \\ 
$R_P$ [$M_J$] & $0.346$ & - & $3.7_{-3.4}^{+22.2}$ & 
$3.2_{-3.1}^{+82.2}$ \\
$M_S$ [$M_{\oplus}$] & $1.00$ & - & $19.4_{-19.4}^{+3644}$ & 
$76_{-76}^{+82460}$ \\
$R_S$ [$R_{\oplus}$] & $1.00$ & - & $1.4_{-1.3}^{+15.6}$ & 
$1.7_{-1.6}^{+69.3}$ \\
$\rho_S$ [g\,cm$^{-3}$] & $5.5$ & - & $18.5_{-18.4}^{+576.1}$ & 
$79_{-78}^{+5456}$ \\ [1ex]
\hline\hline 
\end{tabular}
\label{tab:NepM2none} 
\end{table*}

\subsection{General Observations}

The simulations above are all for M2 stars with habitable-zone Neptune-like 
planets and Earth-like moons. These parameters were chosen as 
\citet{kipping:2009c} have shown that such cases are optimally detectable for
habitable-zone scenarios. We also tried the same configurations but using a K5 
dwarf ($M_* = 0.67$\,$M_{\odot}$ and $R_* = 0.72$\,$R_{\odot}$) but the longer
period of the habitable-zone (half as many transits in the same time window)
combined with lower radius and mass ratios for both the planet and moon meant
we were unable to find convergent fits. This is not surprising and echoes
the motif of the \emph{MEarth} project \citep{irwin:2009} and the predictions in 
\citet{kipping:2009c}.

In reviewing our fits, we find that for quantities such as $M_S/M_P$ and $s$, 
which are positive definite, an overestimation of their value is common due to 
the boundary condition that they are greater than zero and the generally low
signal-to-noise. This is similar to the situation for orbital eccentricity in
radial velocity fits \citep{lucy:1971}.

\section{Discussion \& Conclusions}
\label{sec:conclusions}

\subsection{Comparison to Previous Exomoon Light Curve Simulators}

Other works in the scientific literature have made use of algorithms to simulate
exomoon transit light curves. To our knowledge, there exists several previous 
uses of such routines. Figure~4 of \citet{sartoretti:1999} presents four 
simulations of planet-moon combined transits, although details on the methods 
use to produce these simulations are sparse. Similarly, \citet{szabo:2006} and 
\citet{simon:2009} present figures with simulated combined transits, although
the focus is on a custom definition of transit timing in the former and the
effects of the Rossiter-McLaughlin effect in the latter (i.e. few details
on the methods use to generate the light curves are provided). Finally,
\citet{sato:2009,sato:2010} present perhaps the most detailed account of a
method to generate planet-moon transit light curves but specifically
negate the longitude of the ascending node, orbital inclination of the 
barycentre, orbital eccentricity of the moon, limb darkening effects and use
constant velocity approximations to model the planet/moon motion. 

In all cases, there is no reference to the \citet{fewell:2006} solution and we 
therefore assume this was not utilized in those works. Therefore, all of the 
previous models must be at least partially numerical in nature since the 
solution of \citet{fewell:2006} is the first instance of an analytic solution 
for the area of common overlap between three circles. In contrast, \luna\ is 
fully analytic and we indeed find no appreciable computational time difference 
between using \luna\ versus the usual \citet{mandel:2002} routine.

\subsection{Distinguishing Mutual Events from Starspot Crossings}

Visual examination of the simulations presented in \S\ref{sec:examples}, in 
particular Figure~\ref{fig:NepM2close_big}, reveals that mutual events (i.e.
when the planet and moon eclipse one another, during the stellar eclipse)
bear a close resemblance to the morphology of a starspot crossing (i.e. when
a planet eclipses a starspot on the surface of the star, during the stellar
eclipse). Examples of starspot crossing events can be found in 
\citet{rabus:2009} and \citet{pont:2007}. This observation leads to the question
as to how one could distinguish between a bona-fide moon and a starspot.

Although a detailed study of this question remains beyond the scope of this 
work, we suggest here that there are several tools at our disposal to make such
a determination. Firstly, starspots are expected to co-rotate with the stellar 
rotation period, $P_{\mathrm{rot}}$, which can be found by, for example, 
tracking the long-term flux variation of the target star (e.g. see 
\citeauthor{henry:2008} \citeyear{henry:2008}). Application of a spot model to 
the light curve can test whether the spots are consistent with a rotation period 
determined through other means \citep{silva:2011}.

A second method we can use is to employ the expressions of 
\citet{weighing:2010}, which make use of the system dynamics to measure the
mass and radius of the host star for exomoon systems. If the determined mass
and radius are inconsistent other methods, such as spectroscopy combined with
stellar evolution modelling, then it is likely the putative exomoon signal is
in fact due to starspots. Whilst a dynamically-determined stellar mass and 
radius consistent with the spectroscopic value is not a proof of an exomoon,
one may evaluate the false-alarm-probability of a starspot coincidentally
inducing features which when modelled with \luna\ cause precisely the correct
stellar mass and radius to be determined. Also, a moon may induce transit timing
effects which give rise a derived lunar density which can be compared to
expectation.

A third method we suggest is that starspot crossing events will vary in 
amplitude when viewed chromatically, whereas the exomoon signal will have
a chromatic variation much lower (due to atmospheric molecular absorption) and 
typically undetectable (unless utilizing a telescope such as JWST, see 
\citeauthor{proc:2009} \citeyear{proc:2009}). As an example, one could observe 
the target simultaneously in the visible and infrared wavelengths to test for 
this chromatic variation.

\subsection{Summary}

In this paper, we have presented a new algorithm called \luna\ for modelling the 
transit light curves of a single planet with a single moon transiting a star.
\luna\ was designed from the outset to satisfy several key criteria:

\begin{itemize}
\item[{\tiny$\blacksquare$}] Analytic (absolutely no numerical components)
\item[{\tiny$\blacksquare$}] Dynamic (inherently accounts for all timing 
effects)
\item[{\tiny$\blacksquare$}] Limb darkening incorporated (including non-linear 
laws)
\item[{\tiny$\blacksquare$}] All orbital elements accounted for (e.g. 
eccentricity, longitude of the ascending node, etc)
\end{itemize}

As a result of being both precise and analytic, \luna\ is a highly potent
weapon in exomoon detection. All of the previously predicted observational
consequences, such as TTV, TDV-V and TDV-TIP, are inherently built into the
routine, plus previously unconsidered effects such as ingress/egress asymmetry.
This is done by modelling the reflex motion of the planet due the moon at every
instance, which is ultimately responsible for all of the aforementioned timing 
effects. Transits and mutual eclipses are accounted for using non-linear limb 
darkening laws meaning \luna\ models all known observational consequences of 
exomoons for transiting systems. Further, physical parameters of the star, 
planet and moon may be derived utilizing the dynamical trick described in 
\citet{weighing:2010}.

We have provided simulations of our new algorithm and re-fitted realistic 
noised data with MCMC methods to demonstrate the inverse retrieval of 
parameters. In one example, these fits demonstrate that the 
\citet{weighing:2010} technique is capable of estimating the stellar mass and 
radius down to $\sim$15\% and $\sim$5\% respectively. For the exomoon this 
becomes $\sim$12\% and $\sim$6\% respectively. Whilst this is only one singular 
realization, the simulations lend credence to the prospect of characterizing the 
internal structure and composition of Earth-mass bodies with current 
instrumentation. We stress that these examples are primarily to illustrate
the effectiveness of our new algorithm rather than estimate the range of
configurations which can be conceivably detected, which will be considered in
future work.

In this work, we have assumed the moon is small ($R_S/R_*\ll1$) but future
work will focus on relaxing this constraint so that gas giant binaries may be
modelled as well.

\section*{Acknowledgments}

D. M. K. has been supported by Smithsonian Institution Restricted Endowment 
Funds, NASA grant NNX08AF23G and UCL, through the Science Technology and 
Facilities Council (STFC) studentships. Special thanks to the anonymous referee 
for their highly useful advise and suggestions.


\appendix

\section{Planet Moon Motion}
\label{sec:motion}

\subsection{Nested Two-Body Model}
\label{sub:nested}

We will here derive the relative sky-projected motions of the planet and the 
moon, as used in the \luna\ algorithm. Our goal is to find analytic expressions
for the terms $S_{P*}$, $S_{S*}$ and $S_{PS}$, which were introduced in
\S\ref{sub:conversion}. We will assume only a single moon in our model
and a single planet. To predict the positions of the planet and moon at any
instant in time, one must solve the three-body problem. Since no general, 
analytic solution exists for this problem, one must choose to work in a 
restricted case where analytic expressions can be employed. 

In this work, we will use the nested two-body model presented in 
\citet{weighing:2010}. This considers the motion of the moon to be independent
of the star in the reference frame of the planet-moon barycentre (dubbed the
``inner frame''). Therefore, in the inner frame the motions are Keplerian. The 
barycentre is then considered to also maintain Keplerian motion around the host 
star in the ``outer frame''. 

As shown in \citet{weighing:2010}, this model is an excellent approximation for 
$a_{SB} \lesssim 0.53 R_H$, where $a_{SB}$ is the moon's semi-major axis around 
the planet-moon barycentre and $R_H$ is the Hill radius of the planet. 
Since all bounded prograde satellites must satisfy $a_{SB} \lesssim 0.4895 R_H$ 
\citep{domingos:2006}, then the nested two-body model encompasses all prograde 
moons. We also point out that one may switch the module computing the motions 
fairly simply, but throughout this paper we will use the nested 
two-body model.

\subsection{Inner Frame}
\label{sub:innerframe}

Let us consider the inner frame first, comprising of the planet and moon 
orbiting a common centre of mass. We define the moon to planet-moon 
barycentre distance, at any instant in time, as $r_{SB}$. The subscript notation
represents S to B or ``satellite'' to ``barycentre'' (where barycentre is 
understood to be the planet-moon barycentre). This notation scheme will be 
employed throughout this paper so that there exists no ambiguity as to what is 
being referred to.

\begin{equation}
r_{SB} = |\mathbf{r}_{SB}| = \frac{a_{SB} (1-e_{SB}^2)}{1+ e_{SB} \cos f_{SB}}
\end{equation}

Where $a_{SB}$ is the satellite to barycentre semi-major axis, $f_{SB}$ is the
true anomaly of the satellite around the barycentre and $e_{SB}$ is eccentricity 
of the satellite's orbit around the barycentre. 

One of the simplest reference frames in which to view the orbital motion is when
the line connecting the apoapse and periapse is the $\hat{x}$-axis and the
orbit lies entirely within the $\hat{x}$-$\hat{y}$ plane. Placing the barycentre 
at the focus lying at $\{a_{SB} e_S,0,0\}$, the Cartesian coordinates of the 
satellite are defined by (lower-case symbols are used to denote this is the 
simple view of the system not yet accounting for an observer's viewing angle):

\begin{eqnarray}
\mathbf{r}_{SB} = \left(\begin{matrix}x_{SB}\cr y_{SB}\cr 
z_{SB}\end{matrix}\right)=
\left(\begin{matrix}r_{SB} \cos f_{SB} \cr r_{SB} \sin f_{SB} \cr 0 
\end{matrix}\right)
\label{eqn:initialSB}
\end{eqnarray}

\begin{figure}
\begin{center}
\includegraphics[width=8.0 cm]{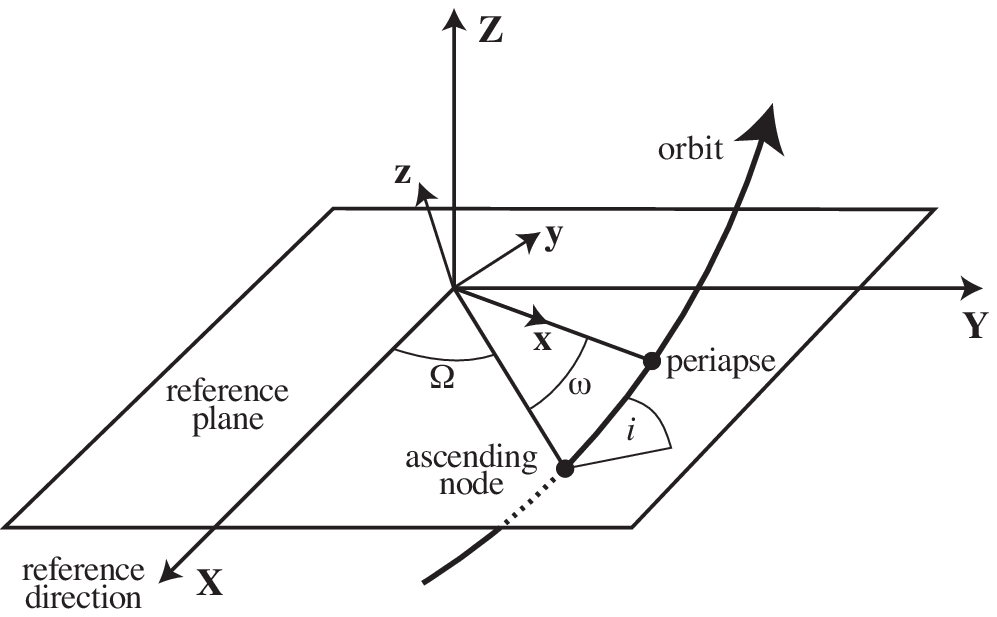}
\caption[Orbital elements of a planet orbiting a star]
{\emph{Orbital elements of a two-body system.
For a transiting planet-star system, $i \simeq 90^{\circ}$ and the 
observer is located at $\{X,Y,Z\} = \{0,0,+\infty\}$. Figure from 
\citet{murray:2010}.}}
\label{fig:complexframe}
\end{center}
\end{figure}

This is highly analogous to the usual equations describing a planet-star system.
In the case of a planet-star, one starts with the analogous version of 
Equation~\ref{eqn:initialSB} and then performs three rotations to account for
the viewing angle an observer has to the system \citep{murray:2010}. These three 
angles are the argument of periapsis, $\omega$, the orbital inclination, $i$, 
and the longitude of the ascending node, $\Omega$. The rotations are performed
sequentially in a clockwise sense for the $\hat{z}$-$\hat{x}$-$\hat{z}$ axes 
respectively and are shown in Figure~\ref{fig:complexframe}. 

The same scheme will be adopted here for the planet-moon inner frame.
For the planet-star case, $i$ is generally close to $\pi/2$ for transits and
thus is a large rotation. However, in our case, we choose to perturb the orbit
from the coplanar condition. Therefore, we define a small angle for
the inclination rotation, $\iota_{SB}$, and use $i_{SB} = \pi/2-\iota_{SB}$ so
that $i_{SB} = \pi/2$ corresponds to a coplanar case, in an analogous way to 
what is used for the planet-star case. The three angles\footnote{$\omega_{SB}$
and $\Omega_{SB}$ take the range $0\rightarrow2\pi$. We also allow $\iota_{SB}$
to take the range $0\rightarrow2\pi$ although only $\pi$ radians are required
to uniquely define every point in space. This is done to prevent boundary
conditions when fitting data.} we need to rotate by are 
$\omega_{SB}$, $\iota_{SB}$ and $\Omega_{SB}$. \citet{murray:2010} showed that 
these rotations may be concisely expressed in matrix notation. Following 
\citet{murray:2010}, we denote the following clockwise rotation matrices about 
the $\hat{x}$ and $\hat{z}$ axes respectively:

\begin{eqnarray}
{\bf P}_X (\phi) =\left(\begin{matrix}
1&0&0\cr
0&\cos\phi&-\sin\phi\cr
0&\sin\phi&\cos\phi\cr \end{matrix}\right)
\label{eqn:Px}
\end{eqnarray}

\begin{eqnarray}
{\bf P}_Z (\phi) =\left(\begin{matrix}
\cos\phi&-\sin\phi&0\cr
\sin\phi&\cos\phi&0\cr
0&0&1\cr\end{matrix}\right)\,.
\label{eqn:Pz}
\end{eqnarray}

Armed with these definitions, the transformation from the simple frame of
$\{x,y,z\}$ to the observed frame of $\{X,Y,Z\}$ may be written 
as\footnote{Note that lower case Cartesian coordinates are used for inner
frames and upper case for outer frames}:

\begin{eqnarray}
\mathbf{R}_{SB} =
{\bf P}_Z (\Omega_{SB}) {\bf P}_X (\iota_{SB}) {\bf P}_Z (\omega_{SB}) 
\mathbf{r}_{SB}
\label{eqn:SBrotn}
\end{eqnarray}

After simplification, this yields the following final Cartesian coordinates:

\begin{align}
X_{SB} &= r_{SB} [ \cos\Omega_{SB}\cos(\omega_{SB}+f_{SB}) \nonumber\\ 
\qquad& - \sin i_{SB} \sin\Omega_{SB}\sin(\omega_{SB}+f_{SB})] \nonumber \\
Y_{SB} &= r_{SB} [ \sin\Omega_{SB}\cos(\omega_{SB}+f_{SB}) \nonumber \\
\qquad& + \sin i_{SB}\cos\Omega_{SB}\sin(\omega_{SB}+f_{SB})] \nonumber \\
Z_{SB} &= r_{SB} \cos i_{SB} \sin(\omega_{SB}+f_{SB})
\label{eqn:finalSB}
\end{align}

Note that $R_{SB} = |\mathbf{R}_{SB}| = |\mathbf{r}_{SB}| = r_{SB}$.
The planet's reflex motion can be found by simply using:

\begin{equation}
\mathbf{R}_{PB} M_P + \mathbf{R}_{SB} M_S = 0
\end{equation}

\subsection{Outer Frame}

To obtain the overall motion, we now need to account for motion of the
planet-moon barycentre around the star. Placing the star at the origin, the
barycentre-star separation, at any instant, is given by:

\begin{equation}
r_{B*} = |\mathbf{r}_{B*}| = \frac{a_{B*} (1-e_{B*}^2)}{1+ e_{B*} \cos f_{B*}}
\end{equation}

Where $a_{B*}$ is the barycentre to star semi-major axis (usually just
dubbed $a$), $f_{B*}$ is the true anomaly of the barycentre around the 
star (usually just dubbed $f$) and $e_{B*}$ is eccentricity of the barycentre's 
orbit around the star (usually just dubbed $e$).

In exactly the same way as was done in the inner frame, the barycentre's
position can be described in a simple frame which is then rotated to
account for the viewing angle of the observer. The simple frame has
the barycentre at Cartesian coordinates:

\begin{eqnarray}
\mathbf{r}_{B*} = \left(\begin{matrix}x_{B*}\cr y_{B*}\cr 
z_{B*}\end{matrix}\right)=
\left(\begin{matrix}r_{B*} \cos f_{B*} \cr r_{B*} \sin f_{B*} \cr 0 
\end{matrix}\right)
\label{eqn:initialB*}
\end{eqnarray}

This is then rotated in the same way as before:

\begin{eqnarray}
\mathbf{R}_{B*} =
{\bf P}_Z (\Omega_{B*}) {\bf P}_X (i_{B*}) {\bf P}_Z (\omega_{B*}) 
\mathbf{r}_{B*}
\label{eqn:B*rotn}
\end{eqnarray}

The last rotation, about $\hat{z}$ by $\Omega_{B*}$, has no bearing on the
transit because the light curve is defined by the star-planet separation only,
which is invariant about a $\hat{z}$-axis rotation since the observer lies at
$z=+\infty$. For this reason, there is no need to include this final angle in 
practice:

\begin{eqnarray}
\mathbf{R}_{B*} =
{\bf P}_X (i_{B*}) {\bf P}_Z (\omega_{B*}) \mathbf{r}_{B*}
\label{eqn:B*rotn2}
\end{eqnarray}

To compute the Cartesian coordinates of the planet and moon (rather than the
planet-moon barycentre), we need to combine the results of the inner frame
derivation with the process of the outer frame. This is achieved by considering
that:

\begin{align}
\mathbf{r}_{S*} &= \mathbf{R}_{SB} + \mathbf{r}_{B*} \\
\mathbf{r}_{P*} &= \mathbf{R}_{PB} + \mathbf{r}_{B*}
\end{align}

The conversion from $\mathbf{r}_{S*}\rightarrow \mathbf{R}_{S*}$ is performed
by repeating the rotations used to go from 
$\mathbf{r}_{B*}\rightarrow \mathbf{R}_{B*}$ (and similarly for 
$\mathbf{r}_{P*}$). This gives:

\begin{eqnarray}
\mathbf{R}_{S*} =
{\bf P}_Z (\Omega_{B*}) {\bf P}_X (i_{B*}) {\bf P}_Z (\omega_{B*}) 
(\mathbf{R}_{SB} + \mathbf{r}_{B*}) \\
\mathbf{R}_{P*} =
{\bf P}_Z (\Omega_{B*}) {\bf P}_X (i_{B*}) {\bf P}_Z (\omega_{B*}) 
(\mathbf{R}_{PB} + \mathbf{r}_{B*}) 
\label{eqn:finals}
\end{eqnarray}

Evaluating, this gives the satellite's position as:

\begin{align}
X_{S*} &= r_{B*} \cos (f_{B*}+\omega_{B*}) \nonumber \\
\qquad& + r_{SB} \cos(f_{SB}+\omega_{SB}) \cos(\omega_{B*}+\Omega_{SB}) 
\nonumber \\
\qquad& - r_{SB} \sin i_{SB} \sin(f_{SB}+\omega_{SB}) 
\sin(\omega_{B*}+\Omega_{SB}) \\
Y_{S*} &= r_{B*}\cos i_{B*} \sin(f_{B*}+\omega_{B*}) \nonumber \\
\qquad& -r_{SB}\sin i_{B*} \cos i_{SB} \sin(f_{SB}+\omega_{SB}) \nonumber \\
\qquad& +r_{SB}\cos i_{B*} \sin i_{SB} \cos(\omega_{B*}+\Omega_{SB}) 
\sin(f_{SB}+\omega_{SB}) \nonumber \\
\qquad& +r_{SB}\cos i_{B*} \sin(\omega_{B*}+\Omega_{SB}) 
\cos(f_{SB}+\omega_{SB}) \\
Z_{S*} &= r_{B*}\sin i_{B*} \sin(f_{B*}+\omega_{B*}) \nonumber \\
\qquad& + r_{SB}\cos i_{B*} \cos i_{SB} \sin(f_{SB}+\omega_{SB}) \nonumber \\
\qquad& + r_{SB}\sin i_{B*} \sin i_{SB} 
\cos(\omega_{B*}+\Omega_{SB})\sin(f_{SB}+\omega_{SB}) \nonumber \\
\qquad& + r_{SB}\sin i_{B*} \sin(\omega_{B*}+\Omega_{SB}) 
\cos(f_{SB}+\omega_{SB})
\label{eqn:finalmoon}
\end{align}

Evaluating for the planet, we find:

\begin{align}
X_{P*} &= r_{B*} \cos (f_{B*}+\omega_{B*}) \nonumber \\
\qquad& - r_{PB} \cos(f_{SB}+\omega_{SB}) \cos(\omega_{B*}+\Omega_{SB}) 
\nonumber \\
\qquad& + r_{PB} \sin i_{SB} \sin(f_{SB}+\omega_{SB}) 
\sin(\omega_{B*}+\Omega_{SB}) \\
Y_{P*} &= r_{B*}\cos i_{B*} \sin(f_{B*}+\omega_{B*}) \nonumber \\
\qquad& +r_{PB}\sin i_{B*} \cos i_{SB} \sin(f_{SB}+\omega_{SB}) \nonumber \\
\qquad& -r_{PB}\cos i_{B*} \sin i_{SB} \cos(\omega_{B*}+\Omega_{SB}) 
\sin(f_{SB}+\omega_{SB}) \nonumber \\
\qquad& -r_{PB}\cos i_{B*} \sin(\omega_{B*}+\Omega_{SB}) 
\cos(f_{SB}+\omega_{SB}) \\
Z_{P*} &= r_{B*}\sin i_{B*} \sin(f_{B*}+\omega_{B*}) \nonumber \\
\qquad& - r_{PB}\cos i_{B*} \cos i_{SB} \sin(f_{SB}+\omega_{SB}) \nonumber \\
\qquad& - r_{PB}\sin i_{B*} \sin i_{SB} 
\cos(\omega_{B*}+\Omega_{SB})\sin(f_{SB}+\omega_{SB}) \nonumber \\
\qquad& - r_{PB}\sin i_{B*} \sin(\omega_{B*}+\Omega_{SB}) 
\cos(f_{SB}+\omega_{SB})
\label{eqn:finalplanet}
\end{align}

\subsection{Sky-Projected Distances}

The usual planetary transit light curve is completely described by the 
sky-projected distance between the planet and the star. In the same way, the
light curve of a planet with a moon is described by the sky-projected distances,
defined as $S = \sqrt{X^2 + Y^2}/R_*$ (units of the stellar radius are used).
Using $r_{B*}' = r_{B*}/R_*$, $r_{SB}' = r_{SB}/R_*$ and $r_{PB}' = r_{PB}/R_*$,
the sky-projected moon-star distance is:

\begin{align}
S_{S*}^2 &= \Big[ r_{B*}' \cos (f_{B*}+\omega_{B*}) + r_{SB}' 
\cos(\omega_{B*}+\Omega_{SB})\cos(f_{SB}+\omega_{SB}) \nonumber \\
\qquad& -r_{SB}'\sin i_{SB}\sin(\omega_{B*}+\Omega_{SB})
\sin(f_{SB}+\omega_{SB})\Big]^2 \nonumber \\
\qquad& + \Big[ r_{B*}' \cos i_{B*} \sin(f_{B*}+\omega_{B*}) \nonumber \\
\qquad& - r_{SB}' \cos i_{SB} \sin i_{B*} \sin(f_{SB}+\omega_{SB}) \nonumber \\
\qquad& + r_{SB}' \cos i_{B*} \sin i_{SB} \cos(\omega_{B*}+\Omega_{SB})
\sin(f_{SB}+\omega_{SB}) \nonumber \\
\qquad& + r_{SB}' \cos i_{B*} \sin(\omega_{B*}+\Omega_{SB})
\cos(f_{SB}+\omega_{SB}) \Big]^2
\label{eqn:Smoon}
\end{align}

And for the planet:

\begin{align}
S_{P*}^2 &= \Big[ r_{B*}' \cos (f_{B*}+\omega_{B*}) - r_{PB}' 
\cos(\omega_{B*}+\Omega_{SB})\cos(f_{SB}+\omega_{SB}) \nonumber \\
\qquad& +r_{PB}'\sin i_{SB}\sin(\omega_{B*}+\Omega_{SB})
\sin(f_{SB}+\omega_{SB})\Big]^2 \nonumber \\
\qquad& + \Big[ r_{B*}' \cos i_{B*} \sin(f_{B*}+\omega_{B*}) \nonumber \\
\qquad& + r_{PB}' \cos i_{SB} \sin_{B*} \sin(f_{SB}+\omega_{SB}) \nonumber \\
\qquad& - r_{PB}' \cos i_{B*} \sin i_{SB} \cos(\omega_{B*}+\Omega_{SB})
\sin(f_{SB}+\omega_{SB}) \nonumber \\
\qquad& - r_{PB}' \cos i_{B*} \sin(\omega_{B*}+\Omega_{SB})
\cos(f_{SB}+\omega_{SB}) \Big]^2
\label{eqn:Splanet}
\end{align}

Another important term is the separation between the planet and moon alone:

\begin{align}
S_{SP}^2 &= \frac{(X_{P*} - X_{S*})^2 + (Y_{P*} - Y_{S*})^2}{R_*^2} \nonumber \\
S_{SP}^2 &= [r_{PB}' + r_{SB}']^2 \Bigg( \Big[ \cos(\omega_{B*}+\Omega_{SB})
\cos(f_{SB}+\omega_{SB}) \nonumber \\
\qquad& -\sin i_{SB}\sin(\omega_{B*}+\Omega_{SB})\sin(f_{SB}+\omega_{SB})\Big]^2 
\nonumber \\
\qquad& + \Big[ \sin i_{B*}\cos i_{SB}\sin(f_{SB}+\omega_{SB}) \nonumber \\
\qquad& - \cos i_{B*} \sin i_{SB} \cos(\omega_{B*}+\Omega_{SB})
\sin(f_{SB}+\omega_{SB}) \nonumber \\
\qquad& - \cos i_{B*} \sin(\omega_{B*}+\Omega_{SB})
\cos(f_{SB}+\omega_{SB}) \Big]^2 \Bigg)
\label{eqn:Smoonplanet}
\end{align}

\subsection{Repeating Units}

Inspection of the above equations reveals several repeating units. Substituting
these units allows for a more simple and intuitive expression of the sky
projected distances. The following two units repeat in all three sky projected
distances:

\begin{align}
\beta &= r_{SB}' \Big[ \cos(\omega_{B*}+\Omega_{SB})\cos(f_{SB}+\omega_{SB}) 
\nonumber \\
\qquad& - \sin i_{SB} \sin(\omega_{B*}+\Omega_{SB})\sin(f_{SB}+\omega_{SB})\Big] 
\\
\gamma &= r_{SB}' \Big[ \cos i_{B*} \sin i_{SB} \cos(\omega_{B*}+\Omega_{SB}) 
\sin(f_{SB}+\omega_{SB}) \nonumber \\
\qquad& + \cos i_{B*} \sin(\omega_{B*}+\Omega_{SB}) \cos(f_{SB}+\omega_{SB}) 
\nonumber \\
\qquad& - \sin i_{B*} \cos i_{SB} \sin(f_{SB}+\omega_{SB}) \Big]
\label{eqn:gammabeta}
\end{align}

Using these substitutions, the sky-projected distances are simplified to:

\begin{align}
S_{S*} &= [r_{B*}' \cos(f_{B*}+\omega_{B*}) + \beta]^2 \nonumber \\
\qquad& + [r_{B*}' \cos i_{B*} \sin(f_{B*}+\omega_{B*}) + \gamma]^2 \\
S_{P*} &= [r_{B*}' \cos(f_{B*}+\omega_{B*}) - (M_S/M_P) \beta]^2 \nonumber \\
\qquad& + [r_{B*}' \cos i_{B*} \sin(f_{B*}+\omega_{B*}) - (M_S/M_P) \gamma]^2 \\
S_{SP} &= [1+(M_S/M_P)]^2 [\beta^2 + \gamma^2]
\end{align}

The substitution terms may be split into terms which vary with time and
those which do not, using a matrix $\mathbf{I}$ defined as the 
``inner'' matrix:

\begin{eqnarray}
  \mathbf{I} = \begin{bmatrix}
    \beta \\
    \gamma
  \end{bmatrix} =
  \begin{bmatrix}
    \beta_{1} & \beta_{2} \\
    \gamma_1 & \gamma_{2}
  \end{bmatrix}
  \begin{bmatrix}
    \sin(f_{SB}+\omega_{SB})  \\
    \cos(f_{SB}+\omega_{SB}) 
  \end{bmatrix}
\label{eqn:superS}
\end{eqnarray}

Where:

\begin{align}
\beta_1 &= -r_s' \sin i_{SB} \sin(\omega_{B*}+\Omega_{SB}) \\
\beta_2 &= r_s' \cos(\omega_{B*}+\Omega_{SB}) \\
\gamma_1 &= r_s' [\cos i_{B*} \sin i_{SB} \cos(\omega_{B*}+\Omega_{SB}) - 
\sin i_{B*} \cos i_{SB}] \\
\gamma_2 &= r_s' \cos i_{B*} \sin(\omega_{B*}+\Omega_{SB}) 
\end{align}

As the matrix form provides a natural way to describe the motions, it is
convenient to extend it to the barycentric motion as well:

\begin{align}
S_{S*}^2    &= [ \mathbf{O}_2 + \mathbf{I}_1 ]^2 + [\mathbf{O}_1 + 
\mathbf{I}_2 ]^2 \\
S_{P*}^2    &= [ \mathbf{O}_2 - (M_S/M_P) \mathbf{I}_1 ]^2 + [ \mathbf{O}_1 - 
(M_S/M_P) \mathbf{I}_2 ]^2
\end{align}

Where:

\begin{align}
\mathbf{O}_1 &= \psi = r_{B*}' \cos i_{B*} \sin(f_{B*} + \omega_{B*}) + 0 
\nonumber\\
\qquad&= \psi_1 \sin(f_{B*}+\omega_{B*}) + \psi_2 \cos(f_{B*} +\omega_{B*}) \\
\mathbf{O}_2 &= \epsilon = 0 + r_{B*}' \cos(f_{B*} + \omega_{B*}) \nonumber \\
\qquad&= \epsilon_1 \sin(f_{B*} +\omega_{B*}) + \epsilon_2 
\cos(f_{B*}+\omega_{B*})
\end{align}

Where we have used:

\begin{align}
\psi_1 &= r_{B*}' \cos i_{B*} \\
\psi_2 &= 0 \\
\epsilon_1 &= 0 \\
\epsilon_2 &= r_{B*}'
\end{align}

Here, $\psi_1 = b_{B*}$, the impact parameter of the barycentre across the
face of the star, and $\epsilon_2 = a_{B*}/R_*$ for a circular orbit. These two 
parameters are typically fitted for in planetary light curve analysis, but it is 
widely known that they share a strong correlation. In matrix form we have the
``outer'' matrix as:

\begin{eqnarray}
 \mathbf{O} = \begin{bmatrix}
    \psi \\
    \epsilon
  \end{bmatrix} =
  \begin{bmatrix}
    \psi_1 & 0 \\
    0 & \epsilon_2
  \end{bmatrix}
  \begin{bmatrix}
    \sin(f_{B*}+\omega_{B*})  \\
    \cos(f_{B*}+\omega_{B*}) 
  \end{bmatrix}
\label{eqn:superP}
\end{eqnarray}


\end{document}